\documentclass[onecolumn,showkeys,reprint,superscriptaddress,nofootinbib,amsmath,amsfonts,amssymb,aps,floatfix,]{revtex4-1}
\usepackage{amsmath,amsfonts,amssymb}
\usepackage{subfigure}
\usepackage{graphics,graphicx}
\usepackage{xcolor}
\usepackage{url} 
\usepackage{hyperref}
\usepackage[a4paper, total={7in, 10in}]{geometry}
\usepackage{xspace}

\begin{document}

\title{Generating surrogate temporal networks from mesoscale building blocks}

\author{Giulia Cencetti}
\affiliation{Aix Marseille Univ, Universit\'e de Toulon, CNRS, CPT, Marseille, 13009, France}
\author{Alain Barrat}
\affiliation{Aix Marseille Univ, Universit\'e de Toulon, CNRS, CPT, Marseille, 13009, France}

\begin{abstract}
Surrogate networks can constitute suitable replacements for real networks, in particular
to study dynamical processes on networks, when only incomplete or limited datasets are available.
As empirical datasets most often present complex features and interplays between structure
and temporal evolution, creating surrogate data is however a challenging task, in particular
for data describing time-resolved interactions between agents. Here we propose 
a method to generate surrogate temporal networks that mimic such observed datasets.
The method is based on a decomposition of the original dataset into small temporal subnetworks encoding local structures on a short time scale.
These are used as building blocks to generate a new synthetic temporal network that will hence inherit the shape of local interactions from the dataset.
Moreover, we also take into account larger scale correlations on structural and temporal dimension, using them to inform the process of assembling the building blocks.
We showcase the method by generating surrogate networks for several datasets of social interactions and comparing them to the original data on two complementary aspects. First, we show that the surrogate data possess complex structural and temporal features similar to the ones of the original data. Second, we simulate several dynamical processes, describing respectively epidemic spread, opinion formation and emergence of norms in a population, and compare the outcome of these processes on the generated and original datasets.
We describe the method in detail and provide an implementation so that it can be easily 
used  in
future works based on temporally evolving networks.

\end{abstract}
\maketitle

\section{Introduction}

Networks are a reference representation tool in the field of complex systems, well suited to describe systems composed of multiple interacting agents~\cite{barrat2008dynamical,posfai2016network,latora2017complex,newman2018networks}. 
Formed by a set of discrete nodes and the connections between them, networks (or graphs) schematize the existing interactions among elements, providing a representative picture of the system architecture in many disciplines, from physics to sociology, biology, and economy. 
In many cases, agents’ interactions undergo a temporal evolution, with links appearing and disappearing over time, and their description necessitates the use of temporal networks representations and tools~\cite{holme2012temporal, masuda2016guide}. 
This framework proves valuable in many settings, like neuronal functions and ecosystems, but is particularly useful to describe social contexts, where connections among people change over time, both in physical and remote interactions, with non-trivial temporal and structural correlations. 
Both static and temporal network representations allow moreover to describe the unfolding of dynamical processes among agents such as, for instance, spreading phenomena (of diseases, opinions or information), transportation models, communication, synchronization, and consensus formation \cite{barrat2008dynamical}. 
The collective behaviors resulting from these processes depend on both the structure of the underlying network and its temporal evolution~\cite{barrat2008dynamical,castellano2009statistical,pastor2015epidemic,
holme2012temporal}, so that the correct description of their properties requires their study
and simulation on temporal networks with realistic structural and temporal patterns.


However, information about real temporal sequences of interactions between agents is often incomplete, limited in size and duration, due to the many challenges in collecting datasets
\cite{holme2012temporal,eames2015six,genois2015compensating}. It is thus not 
straightforward to study processes on empirical temporal networks of large size and of durations long enough with respect to the processes' timescales. In this context, synthetic networks that mimic the observed complex patterns of real structures can serve as surrogate substrates on which to simulate processes.
Generating synthetic temporal networks with realistic features and arbitrary sizes and timescales is however 
a challenging task: 
the interplay between temporal evolution (e.g., how each connection changes in time, the durations of interactions, the frequency at which new ones appear and existing ones disappear) and topological organization 
(the structural correlations between the instantaneous relations among nodes)
leads indeed to convoluted spatio-temporal structures that require specifically designed tools 
to analyze and are difficult to reproduce
\cite{braha2006centrality,isella2011s,masuda2019detecting,galimberti2020span,pedreschi2022temporal,hartle2024autocorrelation,karsai2012universal}.


The methods for temporal network generation can be broadly divided into two categories: the first is represented by theoretical models~\cite{castellano2009statistical}, which start from
governing rules assumed to be the basis for the evolution 
mechanisms of the agents' interactions. These rules are thus used to 
build connections among nodes and determine their temporal evolution. 
Usually, these models are not targeted at mimicking a specific observed network but aim at investigating and validating the possible underlying mechanisms of temporal network
dynamics, by trying to recreate some general features considered as relevant.
A well-known example is constituted by the activity driven models~\cite{perra2012activity,laurent2015calls,starnini2013topological,karsai2014time,lebail2023modeling}: in their simplest versions, they focus on reproducing the heterogeneity between nodes' activations, without reproducing complex temporal properties. Refined versions
show that short and long-term memory effects in the activation of nodes and links
are needed to reproduce temporal features such as the heterogeneity of contact durations
and of the times between subsequent contacts, or the emergence of correlated evolution patterns
of groups of nodes (communities)
~\cite{stehle2010dynamical,zhao2011social,lebail2023modeling,moinet2015burstiness,moinet2016aging,kim2018dynamic,williams2022shape,hartle2024autocorrelation,
rocha2013bursts,vestergaard2014memory,jo2011emergence,laurent2015calls}.
While this category of models relies thus on a priori imposed theoretical inductions
to produce desired outcomes,
the second category is represented by emulative algorithms for surrogate networks~\cite{genois2015compensating,purohit2018temporal,zhou2020data,presigny2021building,zeno2021dymond,duval2024algorithm,calmon2024preserving}. 
In these somewhat more applicative procedures, one starts from a specific input network and
tries to create a surrogate network that mimics the input structure, i.e., that
produces patterns similar to the ones observed in the input, without making
any theoretical assumption about the underlying mechanisms
(Machine Learning processes can be used for this goal~\cite{zhou2020data}, too). 

Here, we propose an alternative methodology to generate surrogate temporal networks, which borrows aspects from both approaches. On the one hand, the method 
can be ascribed to the category of emulative algorithms, as it aims at producing 
surrogates of given empirical temporal network datasets. On the other hand, it also
leverages several theoretical assumptions on the mechanisms for the establishment and dynamics
of links. First, similarly to models in which the activity of nodes depends on their recent
prior interactions \cite{stehle2010dynamical,zhao2011social,gelardi2021temporal,lebail2023modeling}, it 
focuses on the behavior of individual nodes, on their recent past and on their local neighborhood
to determine their future interactions. 
Such approach was developed by Longa et al. \cite{longa2024generating} and was shown to reproduce several time averaged quantities (such as the number of instantaneous neighbors of a node, averaged over nodes and time) and 
instantaneous ones (such as the total number of interactions at each time). 
It however fails to reproduce other relevant properties observed in empirical data,
such as the large values of the clustering~\cite{newman2018networks} (i.e. the fraction of completely connected triads of nodes), the organization of nodes into groups~\cite{fortunato2010community,newman2018networks}, and the heterogeneity in the 
overall activity of connections (the total time in interaction of pairs of nodes 
can vary over orders of magnitude \cite{isella2011s,barrat2015face}).
To overcome these limitations, we introduce the second assumption
that some network features observed in real data result from meso-scale topological correlations and long-term temporal correlations, and we thus include 
mechanisms to reproduce global and meso-scale properties observed in the real temporal networks that we want to mimic. 
%
Moreover, this methodology makes it possible to generate surrogate networks with a different temporal extension from the one that it takes as input. It can therefore be used for augmenting data, providing a solution to the 
problem of data with limited duration \cite{calmon2024preserving}.

In the following we first provide a detailed explanation of the methodology. We
use it to produce surrogate datasets of five empirical temporal networks
describing social interactions in different contexts, namely
face-to-face contacts in a primary school~\cite{stehle2011high}, a high school~\cite{fournet2014contact,mastrandrea2015contact}, a workplace~\cite{genois2018can}, and a conference~\cite{isella2011s}. We assess the similarity between the surrogate and original
datasets along a series of structural and temporal characterization tools of temporal networks.
We then consider three dynamical processes with different dynamical properties, to showcase the possibility to use such surrogate data in numerical studies
of these processes: a generic model for the spread of infectious diseases  \cite{pastor2015epidemic,barrat2008dynamical}, the Deffuant model for
opinion dynamics~\cite{deffuant2000mixing,zarei2024bursts}, and the Naming Game model
for the emergence of conventions in a population \cite{baronchelli2007nonequilibrium,baronchelli2016gentle}.
We also show the results obtained with two simplified versions of the method, which allow us
to outline 
the role and the importance of the different mechanisms at play
in the generation of the surrogate data.

\begin{figure*}
\includegraphics[width=1\textwidth]{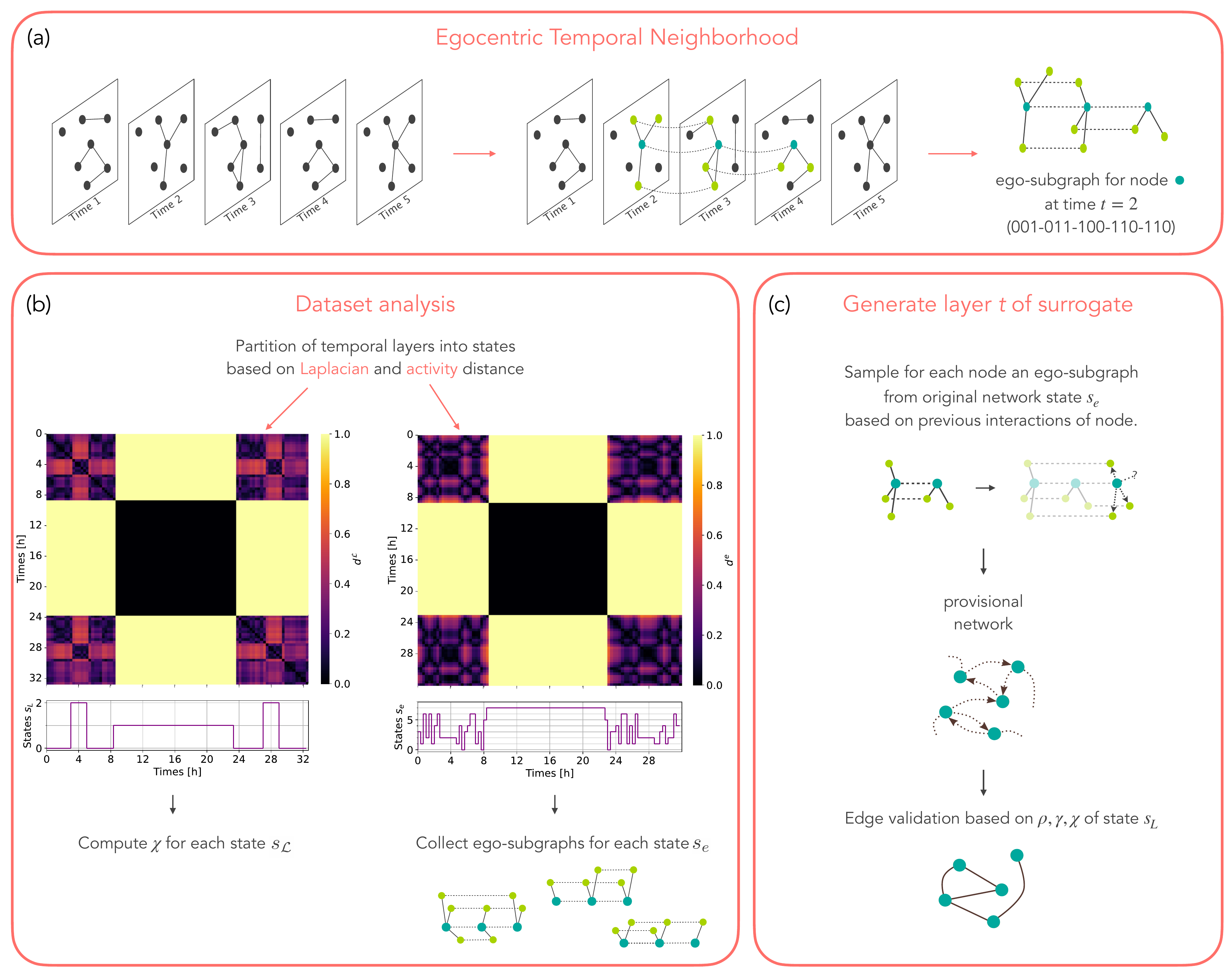}
\label{fig_schema}
\caption{
(a): Example of Egocentric Temporal Neighborhood extraction from a temporal network. An ego-node (in aqua green) and its neighbors (in light green) are highlighted for three time steps ($d=2$) and the corresponding ego-subgraph is depicted on the right together with its univocal signature.
(b): The temporal snapshots are partitioned into states based on $d^{\mathcal{L}}$ (left) and $d^e$ (right). 
The two matrices report the distances among snapshots for a temporal network describing
interactions between children in a primary school \cite{stehle2011high}, with temporal resolution of 5 minutes. For $d^e$ we use  $k_d$ with $d=2$. 
The plot below each matrix depicts the succession of states associated to each snapshot (the state numbers do not carry any specific meaning, they just serve to label the snapshots).
(c): Generating a generic snapshot $t$ of the surrogate involves several steps. First the extraction of ego-subgraph for each node, which leads to the construction of a provisional static directed network. Then the edge validation based on the original network features that we aim to mimic.
}
\end{figure*}

\section{Generating surrogate temporal networks}

Let us consider a temporal network in discrete time, with $N$ nodes and $T$ 
temporal snapshots. There are in principle no constraints on the characteristics of this network, that we denote as the ``original network'' to distinguish it from the synthetic surrogate networks that our method will generate. 
We first explain the procedure to generate a network with the same number of nodes and the same temporal length of the original and we will discuss later how to adapt the methodology to generate longer networks.

\subsection{Building blocks of the original network}

The generation process begins with the extraction of information from the original network.
The first step consists in decomposing the original network into building blocks, called ``Egocentric Temporal Neighborhoods'', which describe how each node interacts with its neighborhood on a short time scale~\cite{longa2022efficient}. 
The Egocentric Temporal Neighborhood of a node $i$ at time $t$ is indeed
defined as the temporal subgraph of the original network composed by $i$, all the nodes that interact with it between $t$ and $t + d$, and their links with $i$. The connections between the 
neighbors of $i$ are thus not included (see top panel of Fig.~\ref{fig_schema}).
The time length of the subnetwork is given by $d+1$, where $d$ is a free parameter of the method. 
In Fig. \ref{fig_schema} and in all the successive figures we use $d=2$, a comparison between different values of $d$ is provided in the Supplementary Information (SI).

By removing the node identities from an Egocentric Temporal Neighborhood, and just keeping the information about (i) which node is the central one (ego) and (ii) which nodes are the same in successive snapshots, one obtains an \textit{ego-subgraph}, which  encodes just the shape of the local interaction. Note that each ego-subgraph can be mapped to a binary string, with the following procedure \cite{longa2022efficient}: for each neighbor of the ego, we create a binary sequence of length $d+1$ with a $1$ each time the neighbor is connected to the ego, and a $0$ when it is not (in the example of Fig. \ref{fig_schema}, there are thus $5$ sequences of length $3$); we then concatenate these sequences (using the lexicographic order to decide in which order
they are concatenated).

We thus extract the Egocentric Temporal Neighborhood and the resulting ego-subgraph for each node
and at each time. Note that, since the node identities are removed in the ego-subgraph, different Egocentric Temporal Neighborhoods (of different nodes and at different times) yield the same ego-subgraph. 
Overall, the set of all the ego-subgraphs and their numbers of occurrences encode the local interactions of the original network, i.e. how the nodes' neighborhoods are formed and evolve on a short time-scale \cite{longa2022efficient}, and it has been shown that adjusting a model's parameters to reproduce the ego-subgraphs relative frequencies allows to reproduce as well several other temporal properties of the data \cite{lebail2023modeling}. 

\subsection{Decomposition in temporal states}

While the global set of ego-subgraphs provides a representation of all the local interactions in the temporal network, it is not suited to describe several 
meso-scale structural features of the temporal snapshots seen as static networks, such 
as the existence of triangles or of densely connected groups of nodes. It does not either 
give direct insights into the temporal variations that can characterize a temporal network at various temporal scales. 
It cannot, for instance, differentiate between periods of high and low activity, or 
between periods in which nodes form separate groups and others in which they all mix together.

To understand the network's temporal organizaton, we resort to the method of \cite{masuda2019detecting} to gather the temporal snapshots of the original networks
into groups of snapshots with similar structure: each group of snapshots then defines a ``state''
of the temporal network.
We first assess the similarity between each pair of snapshots $(t_1,t_2)$ through the Laplacian distance
\begin{equation}
d_{t_1,t_2}^{\mathcal L} =\sqrt{ \frac{\sum_{n=1}^N [\lambda_n(t_1) - \lambda_n(t_2)]^2}{\max \bigl[ \sum_{n=1}^N \lambda_n(t_1)^2,\sum_{n=1}^N \lambda_n(t_2)^2 \bigr] }} \, ,
\end{equation}
where $\lambda_n(t)$ is the $n^{th}$ largest eigenvalue of the Laplacian $\mathcal{L}(t)$ of
snapshot $t$~\cite{newman2018networks,masuda2019detecting}.
The matrix of distances between all pairs of snapshots of an example dataset (contacts recorded between students of an elementary school during two days \cite{stehle2011high}) are shown 
in the left part of Fig.~\ref{fig_schema}(b). 

Using this matrix of distances, we then cluster the temporal snapshots into states \cite{masuda2019detecting}. 
Such optimal classification is found using the Dunn's index~\cite{dunn1973fuzzy} (see the Methods section~\ref{met_dunn}).
For the example dataset considered in Fig. \ref{fig_schema}, three states are found 
(corresponding in the school to the time spans of the lectures, of the lunch breaks and of the night). 
As we have used the Laplacian distance to determine these states, we call them ``Laplacian temporal states'', $s_{\mathcal L}$. They are characterized by a different organization of nodes into communities. This can be shown by building, for each state, 
a weighted static network aggregating all the temporal snapshots of that state
(i.e., where each link is weighted by the number of snapshots in which it appears), and
examining the community structures of these static networks. In the example of the school
dataset of Fig. \ref{fig_schema}, a simple application of the 
Louvain algorithm~\cite{blondel2008fast} yields a 
partition in 9 groups for the state 0 (lectures), which 
approximately reflects the students school classes \cite{stehle2011high}. The resulting
modularity index is high, equal to $0.74$. The partition that is obtained for the state formed by the lunch breaks snapshots consists instead in 5 groups (modularity $0.38$).
To summarize these differences, we compute for each state $s_{\mathcal L}$  an index $\chi$, corresponding to the ratio of inter-group and intra-group links, (see the Methods section~\ref{met_chi} for the precise definition). For the primary school example, $\chi=0.014$ for the state corresponding to lectures and $\chi=0.15$ for the state of the lunch breaks ($\chi=0$ for the night, where no links are present).

We notice however that the number of interactions of singular snapshots can undergo considerable variations inside a state $s_{\mathcal L}$. It is thus also useful to define a more fine-grained temporal state decomposition, based on  nodes activity. To this aim,
we define $\langle k_d(t)\rangle$ as the average number of a node's distinct 
neighbors between $t$ and $t+d$  
(this quantity is proportional to the mean length of all ego-subgraphs at time $t$), and we compute the distance
\begin{equation}
d_{t_1,t_2}^e = \frac{|\langle k_d(t_1)\rangle - \langle k_d(t_2)\rangle|}{\langle k_d(t_1)\rangle + \langle k_d(t_2)\rangle]} \ ,
\end{equation}
between all pairs of times in $[1,T]$. The resulting matrix of distances in shown in  Fig.~\ref{fig_schema} for the primary school dataset. Using this matrix to cluster the snapshots and extract states of the temporal network results here in 
8 states $s_e$. As these states correspond to different levels of activity, and as the
ego-subgraphs encode information on this activity (through the numbers of neighbors of each ego),
we group the ego-subgraphs into separate sets, each corresponding to a state:
for each Egocentric Temporal Neighborhood collected at time $t \in s_e$, the resulting
ego-subgraph is put into the set associated to $s_e$.

Overall, we thus end up with two different partitions of the temporal snapshots: one based on the Laplacian distance between snapshots, which will be used to adjust the overall group structure of the synthetic temporal snapshots created by our methodology, and one based on the variations of activity, which will determine which set of ego-subgraphs is used to create each surrogate temporal snapshot.

\subsection{Reassembling the building blocks to generate a surrogate temporal network}

\subsubsection{Provisional snapshot}

The surrogate network is progressively built one snapshot after the other, starting from time 1 to the final step $T$, with an iterative procedure. Each snapshot $t$ is built making use of information on the previously built snapshots $t-d, \cdots , t-1$, and of the information on the
Laplacian state $s_{\mathcal L}$ and activity state $s_e$ to which the snapshot $t$ of the original network belongs.

Leaving to the Methods section~\ref{met_init} the description of the initial creation of the first $d$ snapshots, we present here the procedure to generate a generic temporal snapshot $t > d$, which
is inspired by the configuration model for static networks~\cite{newman2018networks}.
While, in the configuration model, one assigns to each node a desired degree, and one tries to match the different nodes in order to satisfy these requests, here we will assign to each node $i$ in the snapshot $t$ a desired set of neighbors, combining neighbors of $i$ in the preceding 
snapshots and new neighbors, in order to eventually obtain a network with the same distribution of ego-subgraphs as the original network. 
Then, similarly to the configurational model, connections are built in order to match the requests of the various nodes.

More in details, for each node $i$ in snapshot $t$ to be generated, we consider its neighborhood in the previous $d$ snapshots (in the surrogate being generated): this is an Egocentric Temporal Neighborhood of length $d$, that we map onto an ego-subgraph $g_i$ (still of length $d$). 
We then extract, from the set of ego-subgraphs (of length $d+1$) associated to the state $s_e$ to which $t-d$ belongs, those whose first $d$ time-steps coincide with the ego-subgraph $g_i$.
Note that, although comparing graphs is in general a computationally heavy process \cite{kobler2012graph}, 
the fact that ego-subgraphs have been mapped onto binary strings makes this step efficient.
The resulting subset of ego-subgraphs determines  
all the possible extensions of the neighborhood of node $i$ in snapshot $t$ which are compatible 
with the statistics of ego-subgraphs in $s_e$. We extract one of these possible extensions
at random, with probability proportional to the observed frequency of the corresponding ego-subgraph in $s_e$. The extension determines the desired neighborhood of $i$ in snapshot $t$, which can include nodes already in interaction with $i$ in the previous time snapshots,
and interactions with new nodes. We represent the first case by directed desired links from $i$ to the neighbors, and the second case by stubs, i.e. half links not yet connected to other nodes (question mark in panel (c) of Fig.~\ref{fig_schema}).

Once this process has been applied to every node $i$ we thus obtain for the snapshot $t$ a static
``provisional'' network with directed links and stubs, which encodes the desired neighborhoods of all nodes.

\subsubsection{Link validation}

In order to build the surrogate snapshot at time $t$, we now need to decide which of the 
desired neighborhood directed links of the provisional network to transform into actual
links of the surrogate network, and how to connect the remaining stubs. Indeed, not all
the desired neighborhood links can be satisfied: for instance, if there is a directed link from $i$ to $j$ but not from $j$ to $i$, it means that
node $i$ needs to connect to node $j$ to have its desired Ego Temporal Neighborhood, but
that a link between $i$ and $j$ in the snapshot $t$ would frustrate the desired
neighborhood of $j$ by making its Ego Temporal Neighborhood different from the one extracted 
in the creation of the provisional network (vice-versa, not having the link between $i$ and
$j$ would fit $j$'s neighborhood but frustrate the one of $i$).
To decide which desired neighborhoods to satisfy, we thus perform now the ``validation'' stage of the procedure, to go from the provisional snapshot to the actual surrogate one.
The validation will depend on the Laplacian state $s_{\mathcal{L}}$ corresponding to time $t$ in the original network.


First, we validate each reciprocal request (when both directed links $i\rightarrow j$ and $j\rightarrow i$ exist in the provisional network), transforming them into links of the surrogate snapshot, as the establisment of such links
contributes to creating desired Ego Temporal Neighborhoods of both $i$ and $j$.
All the unidirectional links ($i \rightarrow j$ without reciprocal request) instead represent links that would satisfy one node and frustrate the other one, 
so we validate half of them \cite{longa2024generating}. 
The choice of the ones to validate is made taking into account 
(i) a mechanism of long term memory on the links, 
(ii) the average node clustering coefficient in the original snapshot, and 
(iii) the network's organization into groups at time $t$.
More specifically, the validation process is designed to preferentially validate links: 
\begin{itemize}
\item (i) that already appeared in previous snapshots of the surrogate network corresponding to the state $s_{\mathcal{L}}$, repeating interactions between nodes that have already met in that Laplacian state;
\item (ii) that increase clustering, i.e. closing triangles by 
connecting nodes that have common neighbors;
\item (iii) where the two nodes belong to the same group, according to the nodes partition in the state $s_{\mathcal{L}}$, with a probability that depends on the value $\chi$ of that state (i.e. if the original network has low/high modularity this effect is weak/strong), thus yielding a level of modularity in the generated snapshot similar to the one of the original snapshot.
\end{itemize}

In practice, we compute for each unidirectional link $i \rightarrow j$, the quantity
\begin{equation}
s_{ij} = (1 + \gamma_{ij})(1 + \rho_{ij})
\label{eq_validation}
\end{equation}
with $\gamma_{ij}$ the number of common neighbors of $i$ and $j$ in snapshot $t$ (counting all the possible bidirectional links to validate)
and $\rho$ the memory matrix defined as follows:
the element $\rho_{ij}$ counts the number of past interactions between $i$ and $j$ in the surrogate that took place in snapshots up to $t-1$ corresponding to the same state $s_{\mathcal{L}}$ of time $t$, divided by the maximum value of $\rho$. 
Favoring pairs with larger $\gamma_{ij}$ implies increasing clustering in the snapshot that we are generating, while favoring pairs with large $\rho_{ij}$ corresponds to a
(long-term) memory mechanism (a reinforcement process known to create heterogeneity between the aggregated strength of the connections~\cite{laurent2015calls,lebail2023modeling}).
We thus sort all the pairs in decreasing order of $s_{ij}$.
Moreover we use the information about the groups to which each node belongs and the value $\chi$ of inter-groups versus intra-groups connections (found in the original network for the state $s_{\mathcal{L}}$ corresponding to time $t$) as follows:
we review the list of unidirectional links, 
starting from the pair with the largest $s_{ij}$, and validate each link with probability $1-\chi$ if $i$ and $j$ belong to the same group,  and with probability $\chi$ otherwise.
We stop the process when we have validated the needed
number of pairs (half of the total). If the end of the list is reached before this number is reached, we choose the missing links at random in the list.
We note that this method is designed to reach large values of the clustering, since one validates preferentially the links with large values of the number of common neighbors. It is however possible to modulate this effect and obtain lower values of the clustering by 
multiplying  $\gamma_{ij}$ by a factor $c<1$ in Eq. \ref{eq_validation}. 

Finally, we couple the stubs (deleting one if we have an odd number of them) using a similar process. We first create the list of all the nodes with stubs, each node being repeated a number of times equal to its number of stubs. We choose a random node $i$ from this list, and compute 
$s_{ij}$ for all the other elements $j \neq i$ of the list.
We sort the nodes $j$ in decreasing order of $s_{ij}$ and, as above, starting from the largest value, accept the match between $i$ and $j$'s stubs with probability $1-\chi$ or $\chi$, depending whether $i$ and $j$ 
belong or not to the same group. As soon as a match is accepted, we create the corresponding
link $(i,j)$ in the surrogate snapshot (if not yet existing), and remove one copy of $i$ and $j$ in the list of nodes.
We repeat the process until all the stubs have been coupled.

At the end of the validation stage, we obtain a validated static network that 
becomes the snapshot $t$ of the surrogate network. We iterate to create snapshot $t+1$ and so on, until we reach the final time $T$. 

Since we are using the memory of past interactions to generate each sapshot, we do not consider valid the first $T_{\rho}$ snapshots that are generated, where memory is not significant yet, so we generate $T + T_{\rho}$ snapshots in total and the surrogate network will correspond to just the last $T$. The parameter $T_{\rho}$ is here arbitrary set corresponding to 24 hours.\\

Note that the surrogate generation process is stochastic, as both the choice of the 
desired Ego Temporal Neighborhoods of the nodes and the validation steps include random processes. Using different realization of the process thus allows us to produce as many
different surrogate temporal networks as needed,
all reflecting the same statistical properties of the original data.
Since the procedure takes into account the statistics of the ego-subgraphs (E), the structural correlations (S) like modularity and clustering, and the long term temporal correlations (T), in the following we will refer to it as the EST model. 
We will consider also two simplified versions: the ES model (without temporal correlations, i.e. setting $\rho_{ij}=0$ $\forall i,j$) and the E model (without temporal nor 
structural temporal correlations, i.e. $\rho_{ij}=0$ and $\gamma_{ij}=0$ $\forall i,j$).

\subsection{Temporal extension}

The procedure described above generates surrogate networks with the same temporal length $T$ as the original network. We note however that it is possible to iterate the procedure beyond $T$ and to create as many additional snapshots as desired, and thus to generate
surrogate networks of arbitrary length. To do this, we need however to associate each time
$t > T$ to a pair of states $s_e$ and $s_{\mathcal{L}}$, as the original network does not contain such information. One possibility is to use the set of known states, with (for instance) a periodicity depending on the context (for the primary school data for instance a periodicity of 24 hours can be suitable to reconstruct the general temporal evolution of a successive days).
Such a procedure allows to extend datasets without repeating identical interactions, but mimicking the general behavior of the original 
datasets to generate new synthetic stochastic interactions.

\begin{figure*}
\includegraphics[width=\textwidth]{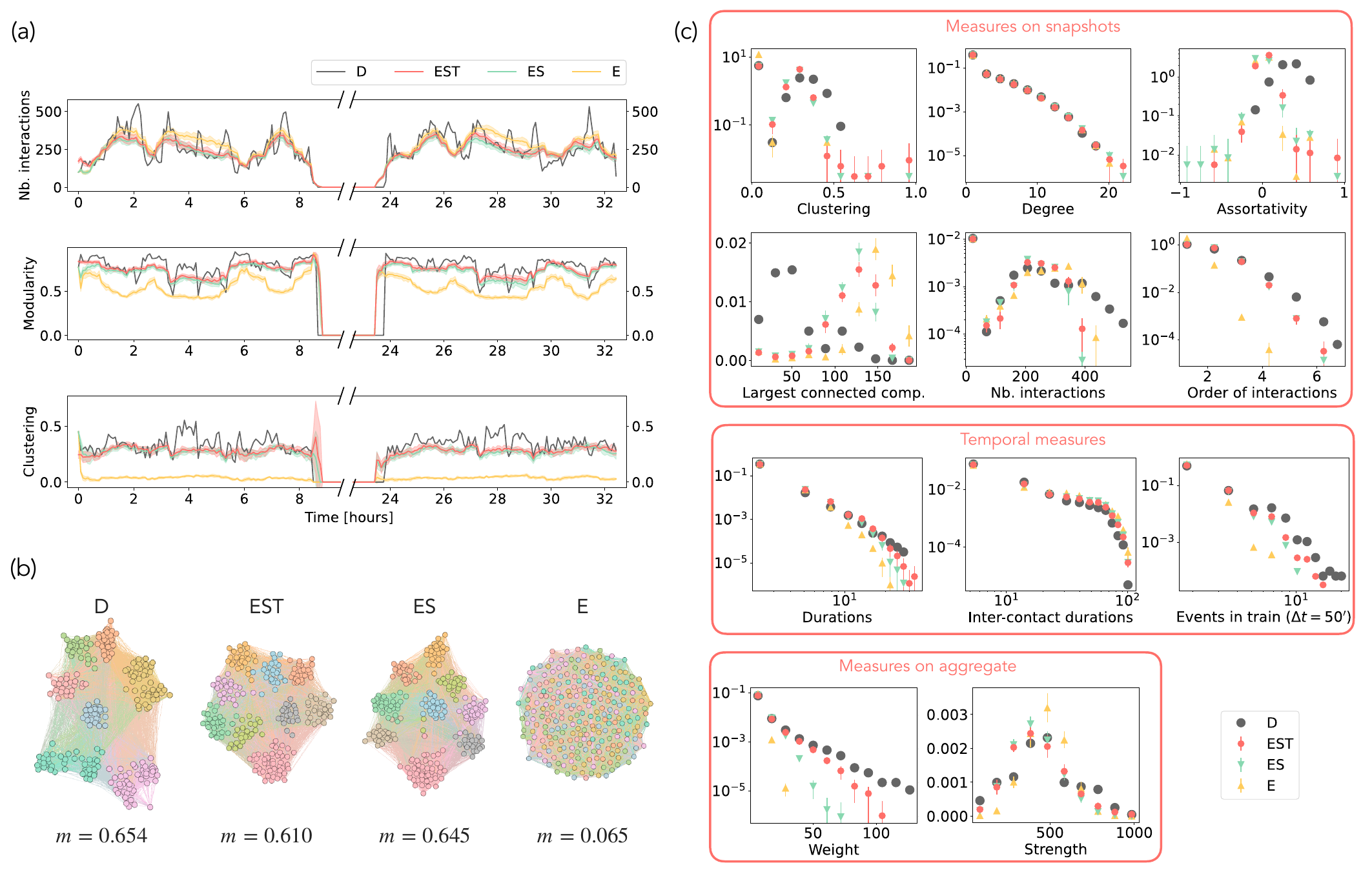}
\caption{(a): Number of interactions, modularity and clustering for each temporal snapshot of original network (D), here the primary school dataset, and for surrogate networks obtained by the EST, ES, and E methods (average over 10 realizations for each method; 
 the shaded area represents the standard deviation).
(b): Aggregated networks for original and surrogates (one realization for each method), where nodes colors reflect the nodes partition obtained by the Louvain algorithm and $m$ is the corresponding modularity index.
(c): Distributions of values for several structural and temporal quantities, divided in three groups according to the measure type. For the surrogate networks we average over 10 realizations
(vertical bars give the standard deviation).
}
\label{fig1}
\end{figure*}

\section{Empirical evaluation}

We illustrate the procedure using 5 temporal networks of social interactions collected by the SocioPatterns collaboration (www.sociopatterns.org) as original networks.
The datasets report face-to-face contacts between sets of anonymized individuals, collected using wearable sensors in diverse environments, including schools, workplaces and conferences,
and have a temporal resolution of 20s.
We show in the main manuscript the results when using the primary school contact data~\cite{stehle2011high} as original network, and show the results for the 
 other datasets in the Supplementary Information (SI). 
 Indeed, this dataset, which describes the interactions between students and teachers during two days, entails rich intertwined structural and temporal features such as groups of nodes (classes) that are densely connected during some periods (classes) and form larger,
 less dense clusters during other periods (lunch breaks).
 
Using this dataset as original network (with here a temporal resolution of 5 minutes) we apply our stochastic procedure (with $d=2$) 10 times, thus obtaining 10 different surrogate networks, for which we report the averaged results. 
Note that we
report the results both for the most refined version of the procedure (EST) and for the two baselines E and ES, in order to understand which aspects of the procedure are most needed to best mimic the various features of the original network. We first compare the surrogate and the original networks under the lens of several structural and temporal features of temporal networks, and we then study whether several dynamical processes unfold in a similar fashion on the surrogate and on the original data.

The basic method that only uses egocentric temporal neighborhoods and ego-subgraphs is very fast (as shown in \cite{longa2024generating} for a method similar to our E process).
It is slightly lowered down by the additional structural and temporal features  but the implementation time remains limited: it takes from 20 seconds to 10 minutes on a standard laptop to generate one surrogate network of those shown in main text and SI with the EST method (the most complicated).
The time variability is mainly due to the number of nodes, that for the networks we tested go from 113 to 327.

\subsection{Structural and temporal properties}

All three procedures (E, ES and EST) produce surrogate temporal networks whose number
of interactions per temporal snapshot mimics well the original one (top panel of Fig.~\ref{fig1}(a)), albeit with smoothed out fluctuations.
This is expected from the method's design, as the original temporal activity has been divided
into states $s_e$, which are reproduced in the procedure.
Similar results were obtained by \cite{longa2024generating}, whose method corresponds to the E procedure with a temporal partition built on an arbitrary time-scale of one hour, instead of being extracted from the data through the states $s_e$.

The lower panels of Fig.~\ref{fig1}(a) report the evolution of more interesting quantities,
namely the average clustering and the modularity of each snapshot. The E model, which does not take into account structural correlations, yields snapshots with very low clustering and lower modularity than the original ones. In contrast, both the ES and EST procedures produce values similar to the ones of the original data.
Panel (b) of Fig.~\ref{fig1} gives an illustration of this point by showing the 
networks resulting from the temporal aggregation of the original and surrogate temporal 
networks (one realization for each surrogate method) where the nodes have been colored according to their partition in communities obtained with the Louvain method \cite{blondel2008fast}. 
The panel also gives the 
value of the modularity $m$ of these aggregated networks, averaged over $10$ realizations of the surrogates (modularity value obtained in each realization through the Louvain method).
Both ES and EST methods yield structures comparable to the original one, while the E method produces a rather homogeneous network with no group structure.

Panel (c) of Fig.~\ref{fig1} gives a more systematic comparison of the properties of surrogates and original dataset by showing the distributions of eleven quantities characterizing the statistics of individual snapshots, temporal statistics and aggregated ones. 

\subsubsection{Instantaneous snapshot properties}

The first six quantities of Fig.~\ref{fig1}(c) are defined for static networks: we compute them on each
single snapshot and display the resulting distributions.
They consist in: clustering, degree, assortativity, size of largest connected component, 
number of interactions and orders of interactions.
The \textit{clustering} is computed as $3\frac{\# triangles}{\# triads}$ 
where $\# triangles$ 
is  the number of closed triangles (triplets $i,j,k$ such that all links $i-j$, $j-k$, $k-i$ exist in the snapshot) and 
$\# triads$ is the number of triads in which at least two of the links exist.
The \textit{degree} refers to the degree of each node in each snapshot (instantaneous 
number of neighbors).
The \textit{assortativity coefficient} measures the correlation between degrees of connected nodes in each temporal snapshot \cite{newman2003mixing}.
The \textit{largest connected component} of a temporal snapshot is the largest subgraph that is connected in that snapshot.
The \textit{number of interactions} refers to the number of links in a snapshot.
The \textit{order of interactions} is defined by promoting all the maximal cliques (groups of all-to-all connected nodes) in each temporal snapshot to higher-order interactions~\cite{battiston2020networks}; the order of an interaction is then given by the number of nodes in a clique minus one (first order interactions involve two nodes, second order involve three, and so on).

In most cases, the distributions computed on the surrogates obtained with the EST procedure are very similar to the original ones. Some propertiees (distributions of degree and of the number of interactions) are actually well reproduced even with the simplest model E. On the other hand,
the E procedure yields small clustering values and small orders of interactions, while taking into account structural correlations in the procedure (Eq. \ref{eq_validation}) yields 
distributions close to the original ones. Note that the generation method considers only pairwise interactions, does not build higher-order interactions (involving more
than two nodes) and does not rely on higher-order concepts. Nevertheless, the 
combination of clustering and community structure effects increases the probability to generate
cliques in the surrogate snapshots, and allows us here to reproduce the statistics of
instantaneous cliques, which can be interpreted in the present context as higher-order
interactions \cite{iacopini2019simplicial,battiston2020networks}.
Two of the distributions are not well reproduced by the surrogates, highlighting some limitations of the model. First, the assortativity is always around zero, independently on the degree correlations in the original network, which are instead slightly positive in all the considered datasets (see SI). In the generation procedure indeed there is no mechanism that takes into account the neighbors' degree when validating the links or connecting the stubs, 
hence no correlation between degrees emerges.
Second, the distribution of the sizes of largest connected components is shifted to larger values with respect to the original network. This is true for all the datasets considered (see SI), and this effect is particularly large for the primary school dataset shown in Fig. \ref{fig1}. The reason can be ascribed to the rigid constraints of a school context, which 
are not directly implemented in the surrogate but only loosely reproduced through the division in temporal states: during lectures at school, the children are separated into classes and there are no interactions between individuals of different classes, so that the largest connected components in the corresponding snapshots are small. In the surrogates instead, this is not strictly implemented: the generation of a group structure, even with a high modularity value, still leaves the possibility for some interactions between groups, which strongly increase the connected components sizes.


\subsubsection{Temporal properties}

The second group of plots in  Fig.~\ref{fig1}(c)  displays the distributions of three important measures characterizing the temporal evolution of interactions in a temporal network, and measured
for all pairs of interacting nodes and over the whole time span of the networks. 
The \textit{duration} of an interaction between two nodes $i$ and $j$ is defined as
the number of consecutive snapshots in which $i$ and $j$ are connected. The
\textit{inter-contact duration} is instead the number of consecutive snapshots between 
two successive interactions of $i$ and $j$ 
(excluding the empty snapshots such as those representing the nights). 
The \textit{burstiness parameter}~\cite{goh2008burstiness,karsai2018bursty} quantifies the heterogeneity of the inter-contact durations distribution as
$B = \frac{\sqrt{n+1}r - \sqrt{n-1}}{(\sqrt{n+1} - 2)r + \sqrt{n-1}}$
with $r$ the ratio between standard deviation and mean inter-contact duration,
and $n$ the sample size (number of inter-contact durations used to create the distribution).
The burstiness is -1 for a periodic  time series, 0 for a Poisson one, and it is positive for a bursty distribution.
Finally, we consider the distribution of the  \textit{number of events in a train}~\cite{karsai2012universal}:
an ``event'' is defined here as an uninterrupted interaction between two nodes (with its duration), and a ``train of events'' as a series of consecutive interactions between the same two nodes such that the interval between the end of an event and the beginning of the successive one is smaller than a parameter $\Delta t$. Multiple trains of events can exist for each pair of nodes and the desired measure is obtained by counting the number of events in each train for each pair of nodes. A broad distribution of this quantity unveils the presence of temporal correlations in the network \cite{karsai2012universal}. 

Figure \ref{fig1}(c) shows that the EST procedure yields surrogate networks able to reproduce well 
the statistics of these three quantities, while the E and ES procedures yield 
narrower distributions of the interaction durations and of the numbers of events in a train.
The burstiness of the EST surrogate (averaged over 10 realizations) is also closer to the original one ($B_{data} = 0.33$,  $\langle B \rangle_{EST} = 0.28$, $\langle B \rangle_{ES} = 0.26$, and $\langle B \rangle_{E} = 0.16$) even if all three methods yield comparable inter-contact durations distributions.
Overall, the EST surrogate thus reproduces well the temporal heterogeneities and correlations observed in the original empirical dataset.

\subsubsection{Aggregated network}

The last two quantities displayed in Fig.~\ref{fig1}(c) are measured on the networks aggregated on their whole temporal span: we show in the bottom panel the distributions of the links 
\textit{weights} and of the node \textit{strengths}: the weight of a link is equal to the
number of snapshots in which the link has been active, while the strength of a node is the sum of the numbers of instantaneous neighbors of this node in each snapshot.
For the E procedure, the links weights display a narrow distribution, reaching only small weight values. This is due to the fact that this procedure does not entail structural nor memory effects, so that the new connections of a node created in the generation and validation of a snapshot are taken at random among all possible ones, and are less repeated than in the ES and EST cases. Taking into account structural correlations leads to a slightly broader
distribution but, as expected from previous studies \cite{laurent2015calls,lebail2023modeling}, 
memory effects are needed to obtain distributions more similar to the original ones.


\begin{figure*}[t]
\includegraphics[width=\textwidth]{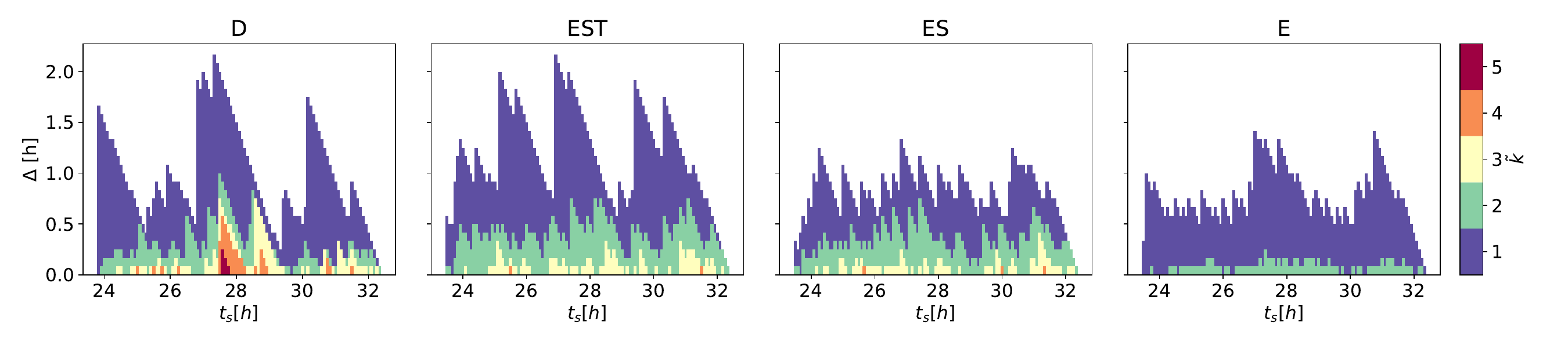}
\caption{Span-core decomposition: for different  $\tilde k$ (different colors) we show the duration of span-cores starting at various times $t_s$ of the second day of the primary school network and of its surrogates.}
\label{fig_span_core}
\end{figure*}

\begin{figure}
\includegraphics[width=0.47\textwidth]{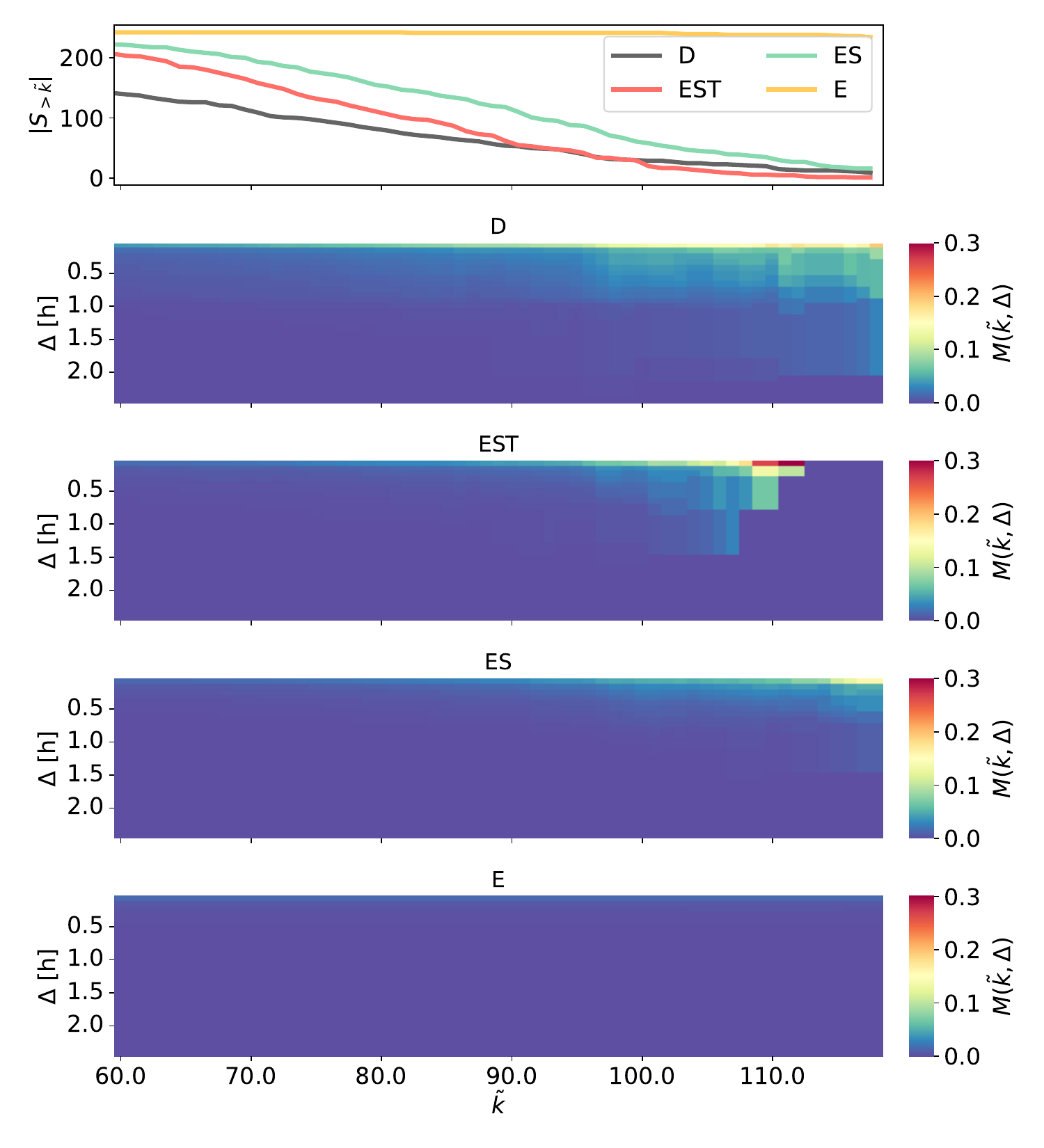}
\caption{Temporal rich club: size of the subset $S_{>\bar k}$ (top panel) and  
maximum $M(\bar k,\Delta)$ over time of $\varepsilon_{>\bar k}(t,\Delta)$, 
as a function of  $\bar k$, for several values of $\Delta$ (lower panels).
}
\label{fig_TRC}
\end{figure}

\subsubsection{Temporal network structures}

Several characterization tools have been developed recently to analyze 
the complex interplay and correlations between topological and temporal aspects of temporal 
networks: we consider here two of these tools, which 
highlight the existence of mesoscale structures in temporal networks.

We first perform the span-core decomposition~\cite{galimberti2020span} of the original and surrogate networks: it decomposes a temporal network into hierarchies of subgraphs of controlled duration and increasing connectivity, generalizing the core-decomposition of static graphs. 
Specifically, a span-core $g$ of order $\tilde k$ is defined on an interval of $\Delta$ consecutive timestamps, such that all nodes in $g$ have at all timestamps of that interval at least $\tilde k$ stable neighbors in $g$ (i.e., the links to these nodes are present during all timestamps of the interval). 
Highly connected (large  $\tilde k$) and stable (large $\Delta$) span-cores have been shown 
to be structures relevant in spreading processes \cite{ciaperoni2020}, and empirical temporal networks are often characterized by the presence of such structures, which are absent from 
random, uncorrelated temporal networks~\cite{galimberti2020span}.
We report in
Fig.~\ref{fig_span_core} the result of the span-core decomposition for the original dataset and for one instance of each surrogate, by showing the duration and the connectivty
of the span-cores starting at various times. Span-cores of durations comparable to the ones
of the original data are obtained with the EST method, while the E and ES methods, which lack  long-term memory or structural effects or both, yield less rich structures.

We then investigate the temporal rich club phenomenon \cite{pedreschi2022temporal}, i.e.,
the tendency of nodes that are well-connected in the aggregated network 
to form structures that are simultaneous and stable  in a temporal network. 
To measure this tendency, we consider, for the original data and for the surrogates,
the subset of nodes $S_{>\bar k}$ of nodes with degree larger or equal to $\bar k$ 
in the aggregated network, and we compute
$\varepsilon_{>\bar k}(t,\Delta)$, defined for each time $t$
as the fraction of links that connect the nodes of $S_{>\bar k}$ in a stable way 
from $t$ to $t+\Delta$. The maximum $M(\bar k,\Delta)$ over $t$ of $\varepsilon_{>\bar k}(t,\Delta)$ (the Temporal Rich Club coefficient \cite{pedreschi2022temporal}), shown in 
Fig.~\ref{fig_TRC} as a function of $\bar k$ and $\Delta$, quantifies whether 
the connections between the nodes highly connected in the aggregated network were simultaneous, dense and stable. Connected and stable structures are observed in the dataset, with 
a temporal rich club effect of increasing density with increasing $\bar k$, and also in the
surrogate produced by the EST method, while the effect is much weaker for the ES case and totally absent in the E model.

These results show that our methodology, when taking into account both structural and temporal 
characteristics of the original network, is able to generate surrogate data that mimics the highly complex structures observed in the empirical temporal networks.

\subsection{Dynamical processes}

As discussed in the introduction, surrogate networks are particularly important as a substitute to empirical data to perform numerical simulations of a variety of dynamical processes. As the outcomes of these processes depend on numerous properties of the network they take place on, 
we investigate here whether the processes simulated on the original and surrogate data
unfold in a similar manner.
We choose three paradigmatic processes that have been largely studied on static and
temporal networks \cite{barrat2008dynamical}:
(i) a model describing the spreading of a schematic disease \cite{pastor2015epidemic}, 
(ii) a model for opinion dynamics with bounded confidence, i.e., such that agents can change opinion by interacting with other agents whose opinion is not too different \cite{deffuant2000mixing}, and
(iii) a model for the emergence of conventions in a population \cite{baronchelli2016gentle}.

\subsubsection{Spreading dynamics}

\begin{figure*}
\includegraphics[width=\textwidth]{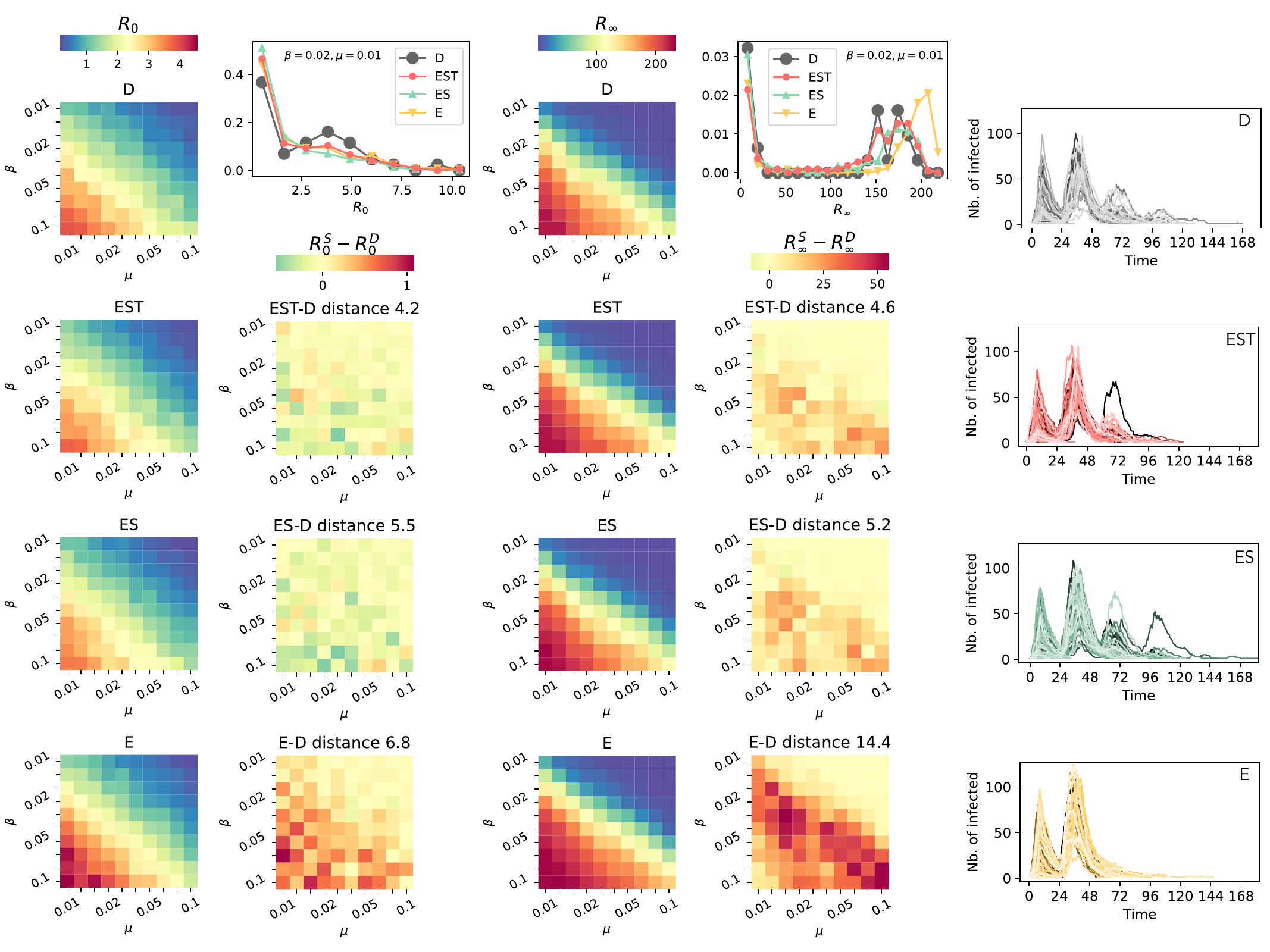}
\caption{
SIR model. 
The heatmaps of the first column show $R_0$ at varying $\beta$ and $\mu$ for original and surrogate networks (averaged over 200 simulations for each pair of parameters). The second column shows the differences between the values obtained by simulations on surrogate networks 
and the ones obtained with the original network.
On top of these heatmaps we report the Canberra distances~\cite{lance1966computer} between the two matrices of values.
We also report on top of the column the distributions of
$R_0$ values for a specific choice of $\beta$ and $\mu$.
Analogously, the heatmaps on the third column show the mean values of $R_{\infty}$ for all the networks and those of the fourth column the differences between surrogates and original network. The plot above reports the distribution of  $R_{\infty}$ for a given pair of parameters $\beta$ and $\mu$.
With the same choice of parameters, the plots on the right show the time evolution of infected individuals on all the networks for 75 simulations. For these specific plots, to facilitate the comparison, the start of the simulations is fixed at $t_0=0$ for all the realizations.
}
\label{fig_SIR}
\end{figure*}

The first dynamics that we consider is the Susceptible-Infected-Recovered (SIR) model for the spread of infectious diseases, where the nodes of the network can only be in one of the
three states: Susceptible nodes (S state) have not yet been reached by the disease, but can be
infected with probability $\beta$ per unit time (per snapshot) when interacting with an infectious node (in I state) \cite{pastor2015epidemic}. 
Infectious nodes then recover spontaneously with probability $\mu$ per unit time, entering the R state. Each simulation starts with one initial seed, 
represented by a node chosen uniformly at random, which is put in the I state at 
a random time $t_0$, while all the others are susceptible. The simulation ends when no
node is in the I state any more (all nodes have either been infected and then recovered, or have remained in the S state). If the last snapshot (time $T$) of the temporal network is reached with the process still active, we repeat the temporal network starting from the first snapshot.

We perfom simulations of the process for varying $\beta$ and $\mu$ and
using the original and surrogate datasets.
For each simulation we compute the basic reproduction number $R_0$ and the final number of recovered $R_{\infty}$, which quantify respectively how many other nodes the initial seed infects, and the final epidemic size, i.e., the size of the network that has been reached by the spread (see the Methods section~\ref{met_sir}). 
The heatmaps in Fig.~\ref{fig_SIR} display the mean values of $R_0$ for the different networks (first column) and the difference between the mean values obtained in simulations on surrogates and dataset (second column), for each set of parameters ($\beta$, $\mu$). 
The third and fourth columns show  heatmaps for the average values of $R_{\infty}$ and for the differences in these average values obtained in the simulations on surrogates and on the original data.
The plots on top of the second and fourth columns show, for a specific choice of $\beta$ and $\mu$, the distributions of $R_0$ and $R_{\infty}$ for each underlying network. Finally, the 
rightmost column shows the temporal evolution of the number of infectious individuals
for $75$ simulations starting all at $t_0=0$ performed on the original network and on each surrogate.

The values, variation patterns and distributions of $R_0$ and $R_{\infty}$ obtained with the EST model are very similar to the ones obtained for simulations run on the original dataset. 
The temporal variations of the number of infectious is also well reproduced. 
A similar picture is obtained when simulations are performed on surrogate networks obtained with the ES model. In surrogates produced by the E model, the lack of structure leads to a faster process with larger impact \cite{pastor2015epidemic}: more nodes are infected at the beginning (larger $R_0$) and the spreading is faster (epidemic curves spanning typically shorter timescales), reaching larger parts of the network (larger $R_{\infty}$). 
In contrast, the group structure present in the ES and EST surrogates has the effect, as in the
original dataset, to limit the possibility of spreading between groups. Moreover, the 
memory effects, which leads to the repetition of interactions, also impact the spread
\cite{calmon2024preserving}. 
Overall, the better reproduction by the EST of the structural and temporal features highlighted
in Fig. \ref{fig1} leads also to outcomes closer to the one of the original data 
\cite{genois2015compensating}.

\subsubsection{Deffuant model for opinion dynamics}


\begin{figure}
\subfigure[]{\includegraphics[width=0.23\textwidth]{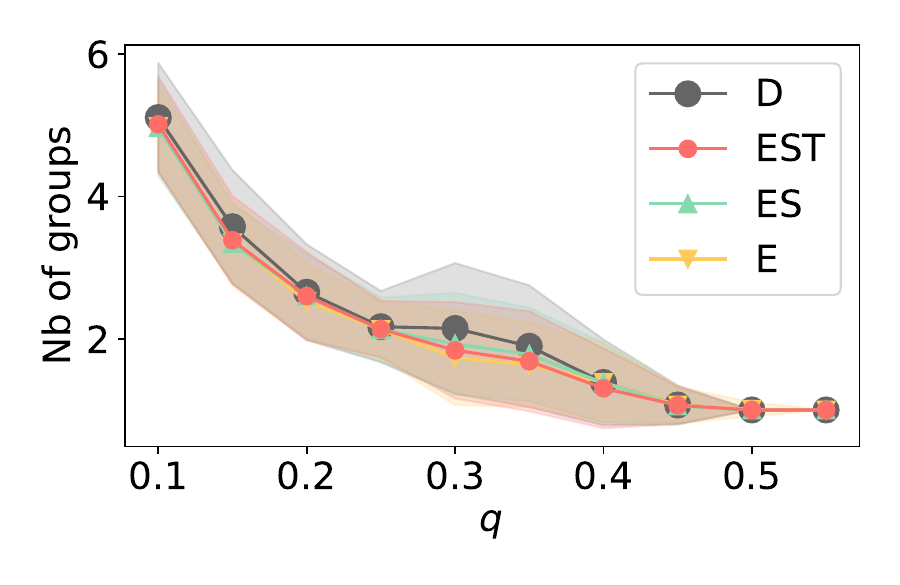}}
\subfigure[]{\includegraphics[width=0.23\textwidth]{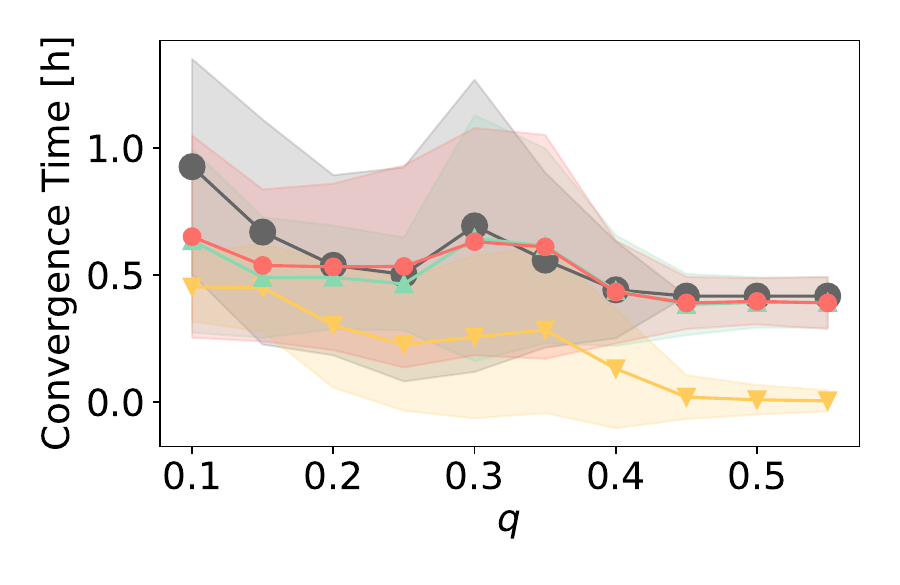}}
\caption{Deffuant opinion model. 
Number of disconnected opinion groups (a) and convergence time (b) at varying $q$, averaged over 200 simulations on each network (for each surrogate method, we perform 10 realizations of the surrogate network
and 20 runs of the opinion model on each realization).
The curves show mean and standard deviation (shaded area).
}
\label{fig_Deffuant}
\end{figure}

The second dynamical process that we consider is the Deffuant model of opinion dynamics \cite{deffuant2000mixing,zarei2024bursts}. 
In this model each node $i$ represents an individual, endowed with an opinion 
represented by a real valued variable $x_i$ between 0 and 1. 
The initial state is given by assigning an opinion extracted uniformly at random 
to each individual. Then, at each snapshot $t$, opinions of pairs of individuals who interact
on the network can change, if and only if their opinions differ less 
than a fixed parameter $q$, i.e. $|x_i(t) - x_j(t)| < q$ (a rule known as
``bounded confidence'', to express the concept that individuals tend to exchange only 
with other individuals whose views do not differ too much). If this condition is met,
the two individuals update their opinion to a common middle ground:
\begin{equation}
\begin{split}
&x_i(t+1) = \frac{1}{2}\bigl(x_i(t) + x_j(t)\bigr)\\
&x_j(t+1) = \frac{1}{2}\bigl(x_i(t) + x_j(t)\bigr) \ .
\end{split}
\end{equation}
The dynamics evolves until the opinions of all interacting pairs cannot change anymore (either because they are already aligned or because they are farther apart than $q$).
See Methods~\ref{met_Deff_NG} for more details.
Such a process tends to align opinions of interacting individuals but does not necessary lead
to a globally aligned population because of the bounded confidence mechanism: 
even in a population where all individuals can potentially interact with each other, 
small values of $q$ lead to the separation into groups of individuals who share
the same opinion, but such that the opinion of different groups differ more than $q$,
making communication between groups impossible \cite{deffuant2000mixing}.

In the case of a process taking place on a (temporal) network, the situation is a bit more complex. The final state corresponds to a separation of the population into groups of nodes inside which all nodes share the same opinion, such that the opinion in a group differs of more than $q$ from the opinion  of the other groups {\it with which it has interactions}. However, there can potentially be two groups of individuals whose opinions differ of less than $q$, but who never interact directly along the network. 
We thus characterize the final state by the number of subnetworks of individuals with 
a homogeneous opinion, which depends on $q$ and on the interplay between the dynamics of opinions and the network's structural and temporal properties.
To obtain this number in practice, we consider the aggregated network, and remove
all the links connecting nodes whose opinions differ more than $q$ in the final state:
we then count the number of remaining connected components (which constitute the opinion subnetworks generated by the process).

We report in Fig.~\ref{fig_Deffuant}(a) the final number of such opinion groups generated
when simulating the Deffuant model on the original and surrogate networks, as a function of $q$.
Panel (b) reports moreover the convergence time of the process, i.e., the number of snapshots after which the opinions do not evolve any more. 
Interestingly, the final number of opinion groups is well reproduced even with the simulations on the surrogate networks obtained with the simplest E method. However, the process is there much faster and the addition of structure in the ES and EST surrogate methods
leads to dynamics much closer to the one on the original network.

\subsubsection{Naming Game}

\begin{figure}
\includegraphics[width=0.5\textwidth]{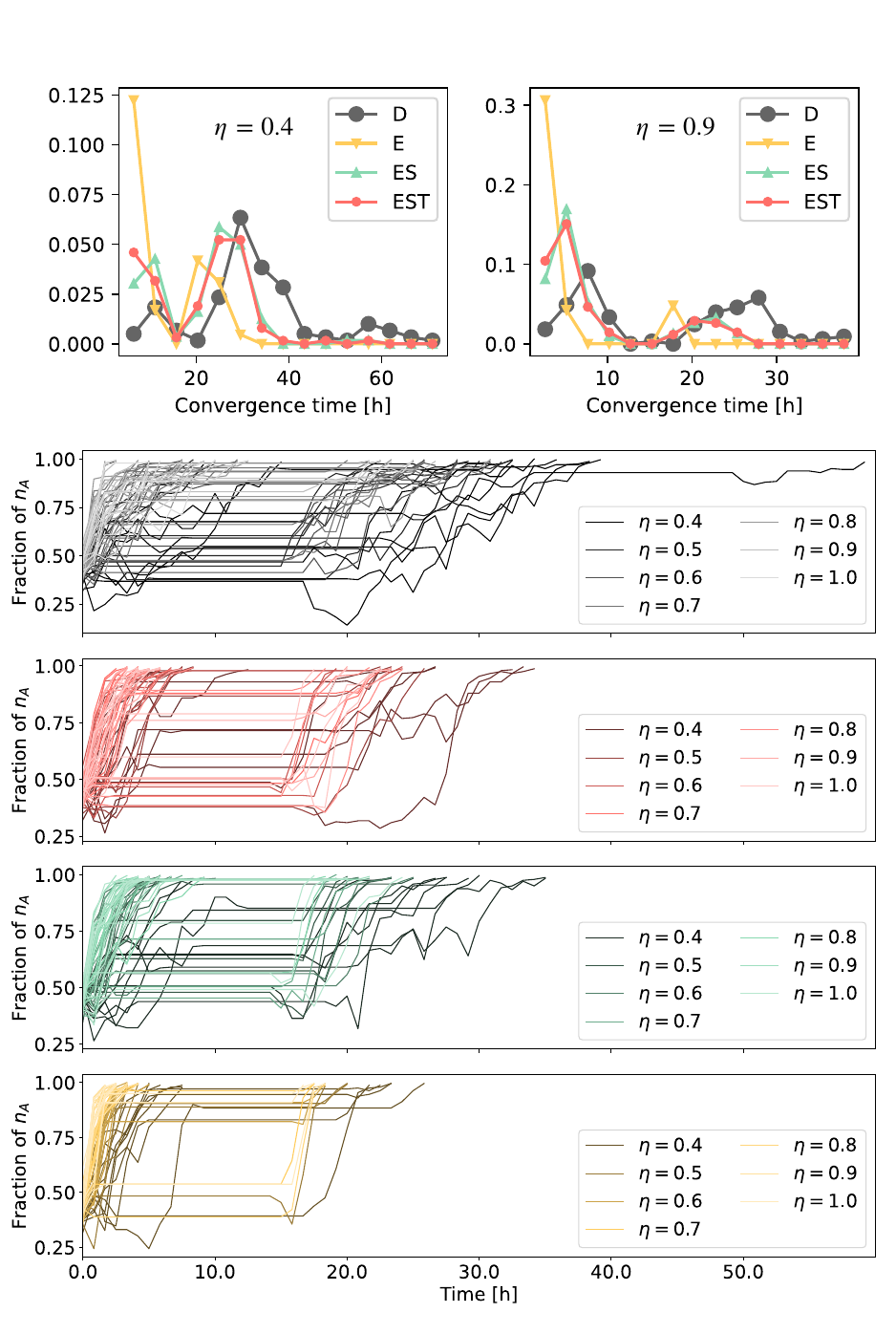}
\caption{Naming game. 
The top panels show the distribution of the convergence time for the Naming Game simulated on all the networks (100 simulations on the original network and 10 simulations on each of the 10 realizations of the surrogate networks) with $\eta=0.4$ and $\eta=0.9$. 
The four plots below show the time evolution of the fraction of nodes that have in their inventory only the name that will eventually win.
For each value of $\eta$ 10 realizations are displayed for each network.}
\label{fig_NG}
\end{figure}

Lastly, we consider the Naming game, a process where the nodes representing individuals 
aim to reach a consensus on the name to give to some object or concept~\cite{baronchelli2007nonequilibrium,baronchelli2016gentle}.
For simplicity we restrict the possibilities to only two names, $A$ and $B$.
Each node has an inventory of possible names which at the beginning contains either $A$ or $B$
(chosen at random).
The simulation starts at a random temporal snapshot of the network. 
At each time, for each interacting pair of nodes, we choose randomly one node to act as
speaker and one as hearer.
The speaker chooses a random name in its inventory and proposes it to the hearer. 
If the hearer does not have it in its inventory, the name is added to the inventory.
If instead the name was already present in the hearer's inventory, the two nodes agree on this name by removing the other possible name. This happens with probability $\eta$, representing the propensity of the hearer to accept the name \cite{baronchelli2007nonequilibrium}. 
The process ends when all the nodes agree on the same name (the other name has then
disappeared from all nodes' inventories). 
See Methods~\ref{met_Deff_NG} for more details.
Figure~\ref{fig_NG} reports the distributions of convergence times for two values of $\eta$ and 
for Naming Game processes simulated on the original data and on surrogate networks. It also
shows, for several realizations of the process, the temporal evolution of 
the number of nodes that have the finally winning name in their inventory.

We observe here a similar phenomenon as in the Deffuant model: simulations performed on the
surrogate network produced by the E method have a typically much faster convergence than 
for the original case. The ES and EST methods produce surrogate networks that better reproduce complex features of the original data, leading to dynamics more similar to the original
one, even if the convergence time remain slightly underevaluated
(note that the peaks in the distribution of convergence times for runs on the primary school 
data correspond to the successive day/night sequence: the process can converge either in the first day, in the second day or in few cases need even an additional simulation day).

\subsection{Robustness}

The Egocentric Temporal Neighborhood strategy is based on the idea that the behavior of a node is strictly related to the immediately previous interactions, where the time scale is set by the parameter $d$.
In the examples that we have shown $d$ was fixed to 2 but values 3 and 4 have been tested and are shown in SI.
The results with $d=3$ are similar to those obtained with $d=2$ (but slightly worse on some important measures like the degree), while those with $d = 4$ turn out to be significantly worse.
In fact, increasing $d$ allows to retain more information about last contacts, but, since many more ego-subgraphs are possible, the frequency of each one is reduced and the results are affected by limited size effect: it is possible that a node in the surrogate explores a set of interactions that does not match any of the ego-subgraphs of the original network. In that case neighbors are progressively removed until a match is found. 
In other words, increasing $d$ can result in overfitting: trying to reproduce ego-subgraphs with more information paradoxically leads to a surrogate whose ego-subgraphs are less conform to those of the original network.

Another robustness investigation is realted to the temporal decomposition into states. As discussed in \cite{masuda2019detecting}, the decomposition can depend on the measure used to 
compute the distance between temporal snapshots, and also on the clustering algorithm
considered (and on the choice of the number of states). 
However, we show in the SI that the properties of the surrogate temporal networks remain similar when using a division between states with an arbitrary time scale of one hour, hinting at a
robustness of the procedure with respect to the choice of the division into temporal states.

\section{Discussion}

The method that we have proposed here to generate surrogate temporal networks starts from the analysis of an original network, supposed to be known and that the surrogate data should
mimic, i.e., the surrogate network, without being identical to the original one, should
have similar statistical properties and structures. This analysis characterizes both
its local and global properties and their temporal evolution. The developed 
methodology leverages the temporal division of the temporal network into states on the one hand, and in collections of ego-subgraphs on the other hand, to generate surrogate 
temporal snapshots one after the other by gluing together different ego-subgraphs, 
implementing in the process a long-term memory mechanism and mimicking the mesoscale
organization of the original snapshots (clustering, organization in groups).
We have shown that the resulting surrogate networks display a complex interplay of structural and temporal properties similar to the one of the original network. Moreover, we have
shown that simulations of a variety of dynamical processes on the surrogate networks yield
outcomes similar to the ones obtained by simulations on the empirical data.

In particular, it is important to note that the three dynamical processes that we have considered represent a broad variety of testing frameworks for our networks, as they differ in several fundamental aspects.
For instance, in disease spreading the nodes states are discrete ($S$, $I$, $R$) and 
each node can only follow an irreversible process from $S$ to $I$ to $R$, without going back.
Nodes having reached the $R$ state do not take part any more in the process.
In the Naming Game model the states are also discrete
($A$, $B$, $AB$) but each node can in principle pass from one state to another one an infinite number of times. In the Deffuant model, the states (opinions) are  
continuous ($x \in [0,1]$) and 
also here there is a priori no limit to the number of times that a node can change opinion.
The global outcomes that result from these process properties are also different and 
highlight different characteristics of the network: in the SIR model a fraction of the nodes is infected and then recover, in the Deffuant model the nodes are partitioned into opinions clusters, and in the Naming Game groups of nodes having converged on one of the names can emerge, before a global convergence is obtained. 
The size of the fraction of the network that is affected by the SIR process,  the number of opinion clusters in the Deffuant model, and the groups created in the Naming Game 
are impacted by the properties of the network of interactions in different ways
\cite{barrat2008dynamical,castellano2009statistical}.
For instance, memory effects impact the Deffuant model and the Naming Game, as 
repeated interactions between a pair of nodes allow them to maintain their agreement 
even if they also interact with other neighbors. In the Naming Game moreover, convergence between a node having only name $A$ and one having only name $B$ requires at least 
two interactions. The SIR model is less impacted by memory, as, once a link has been used
to transmit the infection, it cannot be used for transmission any more.
The SIR process is instead more affected by the burstiness of a temporal network: during
a long time without interaction, a node cannot transmit but can recover spontaneously. On the other hand, its opinion or inventory in the other models simply does not evolve.
Clustering and group structures also favor local spreading and opinion convergence, but
can make it harder for the SIR process to go beyond the group it started from, or for
different groups to converge on a common opinion (this last situation is clearly observed with one of the datasets corresponding to contacts collected
in a high school, where the group structure is very strong, as described in the SI).

The fact that simulations of such various processes on the surrogate networks provide a similar phenomenology
as on the original data highlights thus the versatility of our method. Moreover, 
an important use of surrogate networks consists in providing substrates with realistic properties on which dynamical processes can be studied on various enough time scales, even if long enough datasets are not available. An important example is given by simulations of realistic spreading processes,
which often have longer timescales than most available datasets. 
It is then typically needed to use multiple repetitions of the same temporal network,
which has consequences on the variability of interactions and hence on the realism of the observed behaviors 
\cite{stehle2011simulation,valdano2015predicting,loedy2024repetition,calmon2024preserving}. 
The methodology presented here makes it possible to circumvent this difficulty 
by generating surrogate data of the needed length, avoiding to repeat exactly the same patterns~\cite{calmon2024preserving}. An example is given in SI.

The methodology has also some limitations worth discussing.
First, as the surrogate networks are each based on an original dataset that they mimic, 
a large quantity of information about the data is needed. We have here assumed to have full
knowledge of the original dataset and that it does not suffer from incompleteness.
The case of incomplete data \cite{genois2015compensating}, where the population of nodes is only partially observable and hence the Egocentric Temporal Neighborhoods might be different from those of the complete dataset, is an interesting avenue for future work. 
Second, we have observed that,
even if the method is general and can be applied to any kind of temporal network, we obtain better performances on denser networks. 
In fact the introduction of a preferential link validation by 
Eq. \ref{eq_validation} is more effective if the amount of possible links to choose from at each time step is large enough:  if instead only a few possible links are available, 
the variability is limited and so are the chances to find suitable links that correspond to high values of $s_{ij}$. For temporal networks with too diluted snapshots, we have thus
considered lower temporal resolutions by partially aggregating the empirical data snapshots (e.g.,  on 5 minutes temporal windows for the primary school dataset, 20 minutes for the high schools and the conference, and one hour for the workplace dataset).
%


The methodology we have presented has focused on mimicking the interaction behavior of nodes at short time scales, as well as the instantaneous mesoscale structure
and long-term memory effects. These ingredients have shown to be enough to reproduce a broad range of complex features of temporal networks. The versatility of the method makes it however possible to introduce additional mechanisms to reproduce other properties that could be unveiled by future studies of temporal networks. For instance, the validation stage could be tailored
by changing the definition of $s_{ij}$ to make it depend on other features of the nodes or pairs of nodes (e.g., nodes attributes).
One example of additional mechanism is shown in SI where, instead of the simple modularity, a hierarchical clustering between communities is reproduced in the surrogates.

A further future development of the present work would be to use an existing dataset and produce
realistic surrogate data with larger population sizes. This would yield the important benefit of being able to simulate dynamical processes on networks of large size without needing to actually collect the corresponding data (hence, for instance, without privacy concerns).
Such a development is however far from trivial as the way in which the egocentric neighborhoods change when the population size increases has not yet been investigated, for instance. We therefore plan to tackle this point in future work.

\section{Data and code availability}
The code for generating the temporal networks, to analyze them, and to simulate dynamical processes on them is  available at the following link:
\url{https://github.com/giuliacencetti/Surrogate_net_generation}.
The data of time-evolving social interactions used for the examples are available here: \url{http://www.sociopatterns.org}

\section{Acknowledgments}
This work was supported by the Agence Nationale de la Recherche (ANR) project
DATAREDUX (ANR-19-CE46-0008) to A.B.\\ 
G.C. acknowledges the support of the European Union’s
Horizon research and innovation program
under the Marie Skłodowska-Curie grant agreement No 101103026. 
The funders had no role in study design, data collection and analysis,
decision to publish, or preparation of the manuscript.\\
We acknowledge Antonio Longa and Marco Mancastroppa for the useful discussions.

\section{Methods}

\subsection{Modularity and value $\chi$}
\label{met_chi}

One of the steps of the original network analysis implies to separate the temporal snapshots into states $s_{\mathcal L}$ and to assess their group structure. 
Each state consists of several snapshots, i.e. several static networks.
To find the best partition of nodes into groups we use the Louvain method~\cite{blondel2008fast} on the aggregate network of each state.
Different states will hence correspond to different partitions (for example for the primary school data, the temporal state corresponding to the lectures periods yields a partition
given by the classes, and the partition is different during the lunch breaks).
For each state's static network, we compute the number of links that take place inside a group, $l_{intra}$, and those connecting different groups, $l_{inter}$.
The average density of intra-groups links is then given by
\begin{equation}
p_{intra} = \frac{l_{intra}}{\sum_g \frac{n_g(n_g - 1)}{2}} \ ,
\end{equation}
with $n_g$ the the number of nodes of group $g$, and for the inter-link density we obtain
\begin{equation}
p_{inter} = \frac{l_{inter}}{\sum_g \frac{n_g(N - n_g)}{2}},
\end{equation}
where the denominators correspond to the maximal possible values of $l_{intra}$ and $l_{inter}$, respectively.
We define the parameter $\chi$ as $p_{inter}/p_{intra}$ so it quantifies the probability of having links between groups with respect to inside a group.


\subsection{Generation of the first $d$ snapshots of the surrogate}
\label{met_init}

The first snapshot of the surrogate cannot be based on previous interactions so we build it using the configuration model~\cite{newman2018networks}, which allows us to generate a static network with a given degree distribution, and we choose the degree distribution of the first snapshot of the original network.
To generate the second snapshot we can then rely on the interactions of the first snapshot, so we use the same method described above but using temporal neighborhoods of length $2$ timesteps
instead of $d+1$. We repeat the procedure $d$ times, until we have the $d$ snapshots that we need to initialize the process.


\subsection{Temporal states clustering}
\label{met_dunn}

In the section ``Decomposition in temporal states'', we have briefly described how to decompose the temporal network into 
temporal states, each composed of a set of (non-necessarily contiguous) temporal snapshots.
More in details, once the distances have been computed between all pairs of snapshots, we can cluster them into $C$ clusters by a hierarchical clustering with a bottom up approach: we start by considering each snapshot as a different cluster, and we proceed by iteratively merging the clusters.
We do this for $C \in [3,T]$ for the Laplacian distance and for $C \in [4,T]$ for the activity distance, thus avoiding all the solutions with only a few clusters that would not provide sufficient state variability.
We then chose among all these possible partitions the one that maximizes the Dunn's index~\cite{dunn1973fuzzy}, defined as
\begin{equation}
    \frac{\min_{1\leq c\neq c'\leq C} \min_{i\in c\text{th state}, j\in c'\text{th state}}d(i,j)}{\max_{1\leq c''\leq C} \max_{i', j'\in c''\text{th state}}d(i',j')},
\end{equation}
where $d(i,j)$ is either $d^{\mathcal{L}}$ or $d^{e}$.
The numerator is the smallest distance between two states among all pairs of states and the denominator is the largest distance inside a state among all states.


\subsection{Absent nodes}
\label{met_abs}

The population of a dataset of social interactions can be more or less stable: for some datasets the same set of individuals is present during the entire time span, for other datasets  individuals disappear and new individuals appear making the population vary with time.

The method that we have described to generate surrogate networks does not include such a variability of the population, and in principle all nodes can be active at any time. 
For this reason we include a possible variation, to be applied to datasets where the population change is important.
In fact, if in the original network we observe that some nodes do not participate to the dataset during  specific time intervals, we can reproduce this feature in the surrogate networks too.
This is implemented by imposing that during the same time intervals these nodes, instead of sampling their ego-subgraph from the probability distribution, always choose the empty ego-subgraph, with no interactions.
These nodes will not participate to interactions, hence appearing as absent, in specific intervals of the generated network, reproducing the original network population variability.


\subsection{SIR model: $R_0$ and $R_{\infty}$ }
\label{met_sir}

In the SIR model we compute two  observables that are commonly used to evaluate the disease spreading dynamics.
The first is the basic reproductive number $R_0$, that is numerically computed by counting the number of direct contagions due to the first infected node, i.e. the number of nodes that are infected by this seed node and not by other nodes.
The second observable is the final number of recovered, $R_{\infty}$, that is counted considering only the simulations where at least one contagion event takes place (we exclude all the realizations where $R_{\infty}=1$, which corresponds to runs in which the seed recovered before infecting anybody else).

\subsection{Deffuant model and naming game on temporal networks}
\label{met_Deff_NG}

The Deffuant model of opinion dynamics~\cite{deffuant2000mixing} and the naming game~\cite{baronchelli2007nonequilibrium,baronchelli2016gentle} have been initially formulated for static networks, and can be generalized to temporal ones in several ways~\cite{zarei2024bursts, maity2012opinion}.
In our implementation of both processes, at each snapshot each couple of connected nodes (taken in random order) is considered once and their opinions $x$ or their name inventory are immediately updated (or not).
This means that each node at a generic snapshot can change opinion or change its name inventory as many times as its degree.

For instance, for the Deffuant model, if a node $i$ is connected to node $j$ and $l$ at time $t$, the interaction with $j$ will change the opinion of $i$ from $x_i$ to $x_i'$ and the interaction with $l$ will take place with the updated opinion  $x_i'$, that will again change to  $x_i''$ because of $l$. 
The final opinion of $i$ at time $t$ will be $x_i''$.

The same connections are similarly treated in the naming game: first the roles of speaker and hearer are randomly assigned to $i$ and $j$.
Then their name inventory are updated due to their interaction, and the new name inventory of $i$ will be used in the name exchange with $l$.

The order in which connections are considered at each time step plays a role in the final result. For the sake of generality we choose a random order and repeat the process 100 times to explore a large set of possible evolution patterns.

\bibliography{bib}

\clearpage
\newpage

\appendix

\section*{Supplementary Information for
"Generating surrogate temporal networks from mesoscale building blocks"}
\setcounter{figure}{0}
\setcounter{equation}{0}
\setcounter{table}{0}
\renewcommand{\thefigure}{S\arabic{figure}}
\renewcommand{\thesection}{S\arabic{section}}
\renewcommand{\thesubsection}{S\arabic{section}.\arabic{subsection}}
\renewcommand{\thetable}{S\arabic{table}}
\renewcommand{\theequation}{S\arabic{equation}}

%

\section{Varying the temporal length of ego-subgraphs}

\begin{figure}
\includegraphics[width=0.45\textwidth]{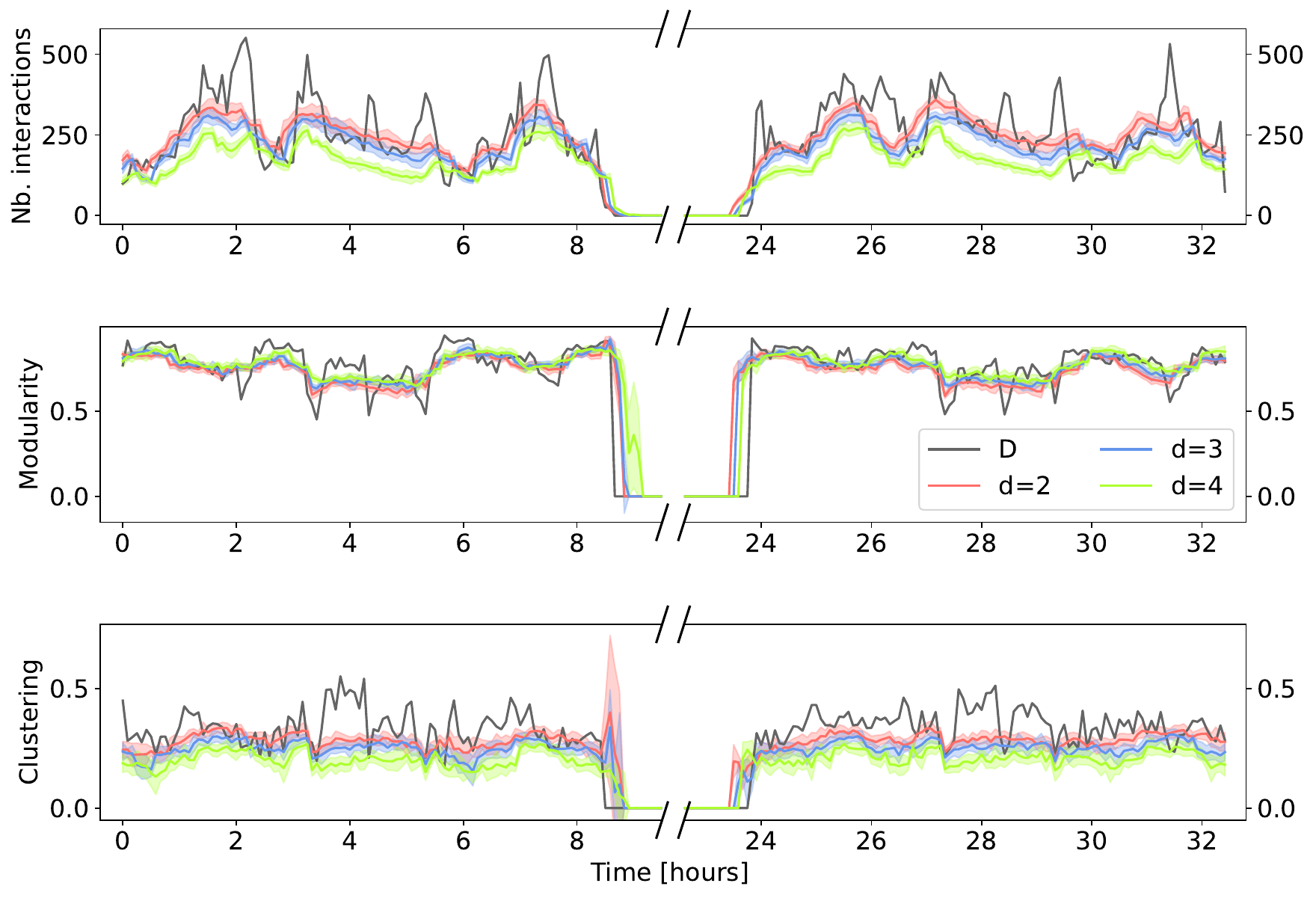} 
\includegraphics[width=0.45\textwidth]{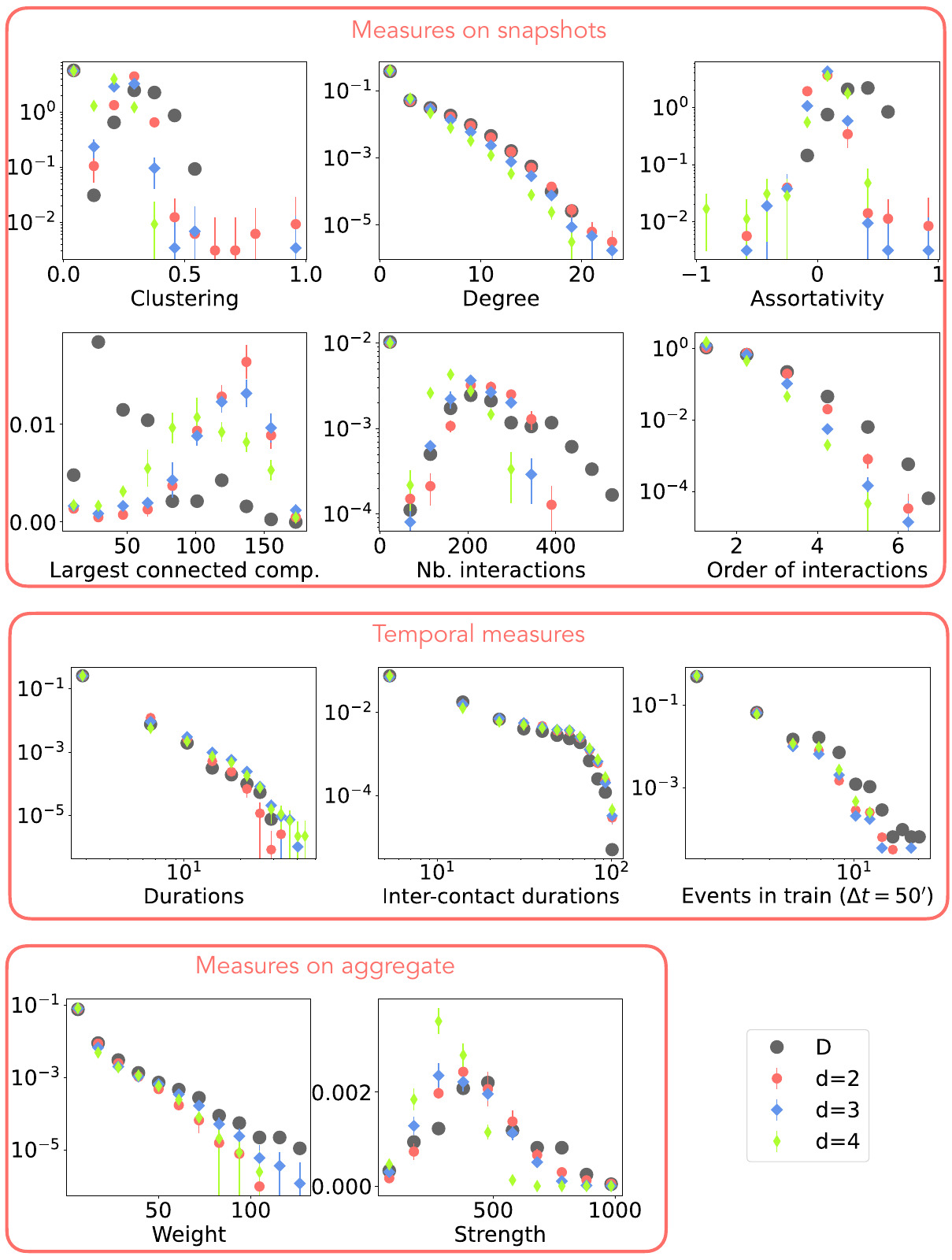}
\caption{\textbf{Testing parameter $d$}. Structural and temporal measures for the primary school dataset (D), and the surrogate networks obtained with the EST procedure at varying $d$.
}
\label{fig_d}
\end{figure}

All the experiments shown in the main text has been performed using ego-subgraphs with three time snapshots, i.e. $d=2$.
However $d$ is a free parameter of our method and it is worth exploring other values.
We show in fig.~\ref{fig_d} the results of experiments performed with $d=3$ and $d=4$ on the primary school dataset.
We observe that if the networks generated with $d=3$ show similar characteristics to those with $d=2$, the quality of the surrogates slightly decreases with $d=4$. 
In fact, increasing $d$ implies considering more information about each node's behavior but it reduces the statistics of the observed ego-subgraphs (the possible ego-subgraphs are more numerous and hence less frequent). 
This increases the probability of not finding a fitting ego-subgraph while generating new interactions, and thus of relying on approximated structures that inevitably imply a generated topology less conform to the original one.


\section{Generating surrogates with hierarchical modularity}
\label{SI_sec_hmod}

\begin{figure}
\subfigure[Original]{\includegraphics[width=0.23\textwidth]{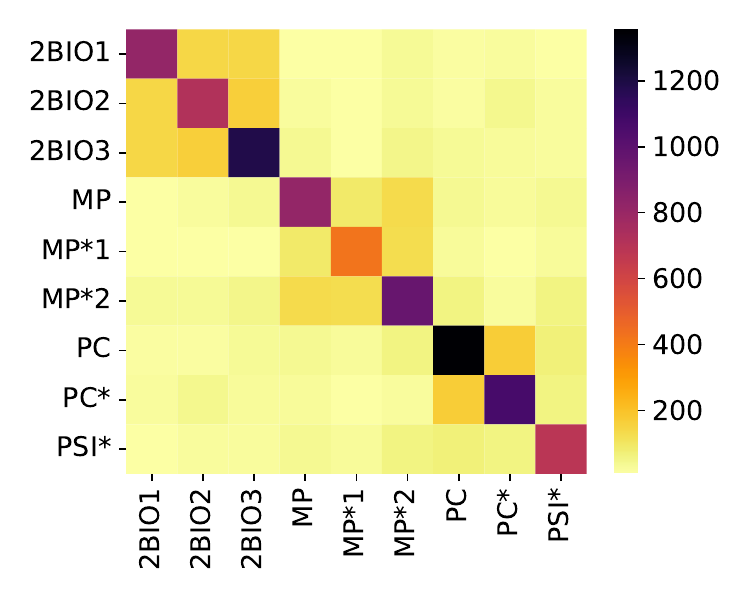}}
\subfigure[Generated]{\includegraphics[width=0.23\textwidth]{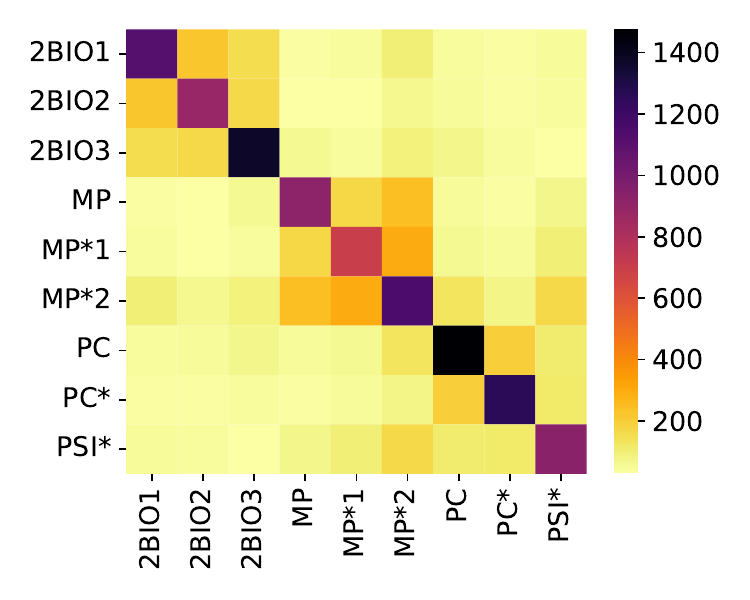}}
\caption{\textbf{Hierarchical modularity}. Contacts among students of different classes in high school 13 dataset~\cite{mastrandrea2015contact} (a) and in surrogate network obtained with hierarchical modularity (b).}
\label{fig_SBM}
\end{figure}

The methodology described in the main text generates temporal networks with a modularity that depends on parameter $\chi$, which measures the ratio between the number of interactions inter and intra communities, where the communities are found using the Louvain algorithm~\cite{blondel2008fast}. 
Some networks are however organized with a hierarchical modularity, i.e. communities of highly connected nodes can be in turn assembled into larger communities, and the larger communites can be  assembled again, and so on, creating a hierarchy of different levels of nodes partitioning.
The nested structure of these networks cannot be captured by the simple subdivision into communities of the Louvain algorithm and this represents a limit of our model.
It is however possible, when dealing with this kind of networks, to implement a slightly different strategy of the proposed algorithm, that we tested on the high school 13 dataset~\cite{mastrandrea2015contact}.
In this high school indeed  students are organized into classes (information given by metadata) and in fig.~\ref{fig_SBM}(a) we report the matrix $L$ which counts the number of contacts inside and among classes measured on the aggregated original network. 
We observe that the majority of contacts takes place intra-classes, however the inter-classes contacts are not equally distributed among couples of classes but form a higher level of communites, in particular a larger community is composed by classes \textit{2BIO1}, \textit{2BIO2}, and \textit{2BIO3}, and another one by \textit{MP}, \textit{MP*1}, and \textit{MP*2}.
Starting from $L$ we define the matrix $P$ of contact probability between all classes (inspired by the Stochastic Block Model algorithm~\cite{holland1983stochastic,karrer2011stochastic}) as:
\begin{equation}
  P_{\alpha\beta} =
  \begin{cases}
    \frac{2 L_{\alpha\alpha}}{n_{\alpha}(n_{\alpha}-1)T} & \text{if $\alpha = \beta$} \\
    \frac{L_{\alpha\beta}}{n_{\alpha}n_{\beta}T} & \text{if $\alpha \neq \beta$},
  \end{cases}
\end{equation}
where $n_{\alpha}$ is the number of nodes in class $\alpha$ and $T$ is the number of snapshots of the temporal network.
This matrix is then used in the link validation of the generation method to replace $\chi$ and $1 - \chi$: all the possible unidirectional links and all the couples of stubs $(i,j)$ are sorted according to $s_{ij}$ (equation (3) of the main text); then, starting from the one with highest $s_{ij}$, each connection is accepted with a probability $P_{\alpha\beta}$, where $\alpha$ and $\beta$ are the indices of the classes to which $i$ and $j$ respectively belong.

By applying this procedure we obtain a surrogate network for the high school 13 whose aggregated contacts are reported in fig.~\ref{fig_SBM}(b), resulting very similar to those of the original network.


\section{Testing temporal networks with extended temporal length}

\begin{figure}
\includegraphics[width=0.5\textwidth]{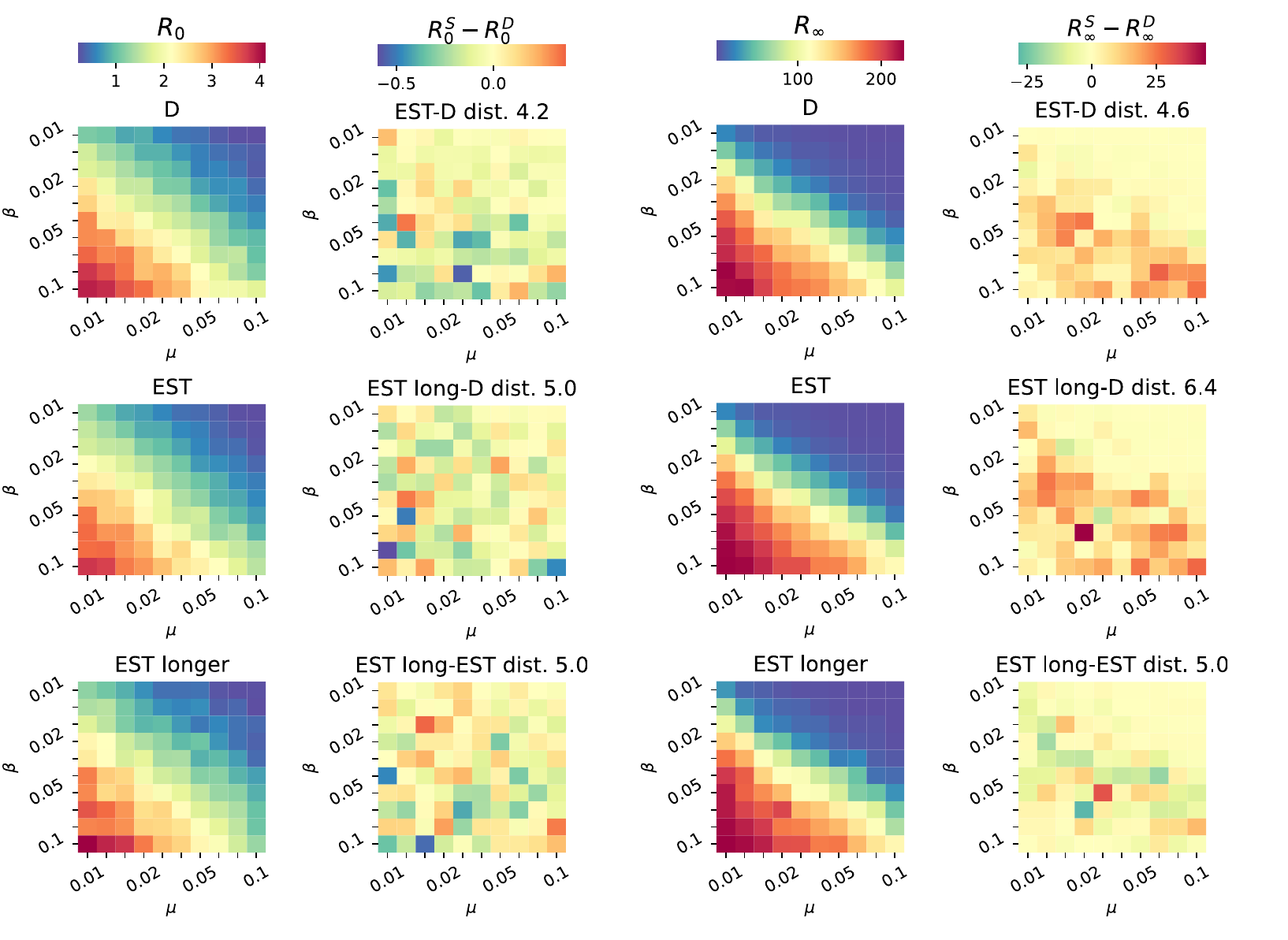}
\caption{\textbf{SIR model with longer temporal length}. The results with longer surrogates (EST longer), lasting 40 days, are compared with those of the EST surrogates of 2 days (EST) and with the primary school dataset (D).}
\label{fig_SIR_longer}
\end{figure}

\begin{figure*}
\subfigure[$\eta=0.4$]{\includegraphics[width=0.24\textwidth]{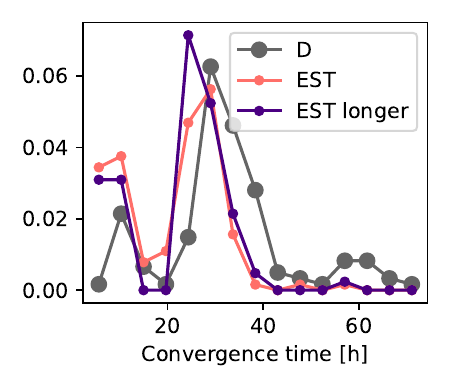}}
\subfigure[$\eta=0.5$]{\includegraphics[width=0.24\textwidth]{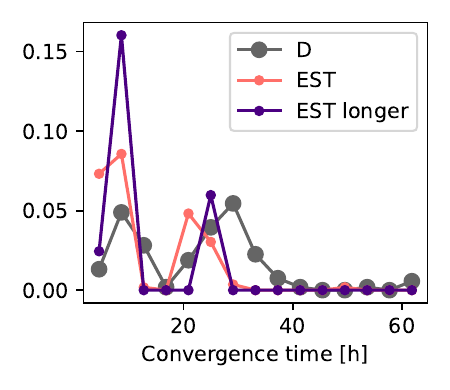}}
\subfigure[$\eta=0.6$]{\includegraphics[width=0.24\textwidth]{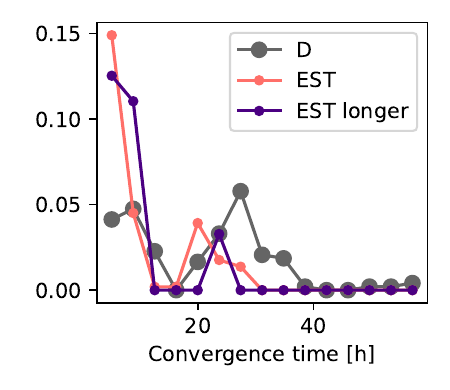}}
\subfigure[$\eta=0.7$]{\includegraphics[width=0.24\textwidth]{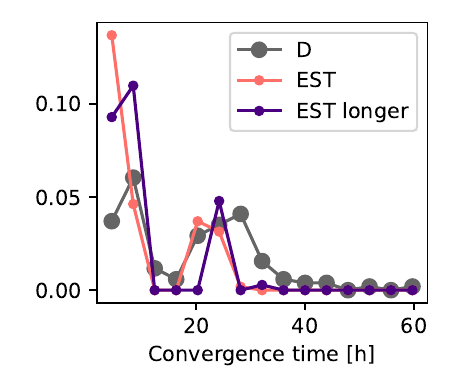}}
\subfigure[$\eta=0.8$]{\includegraphics[width=0.24\textwidth]{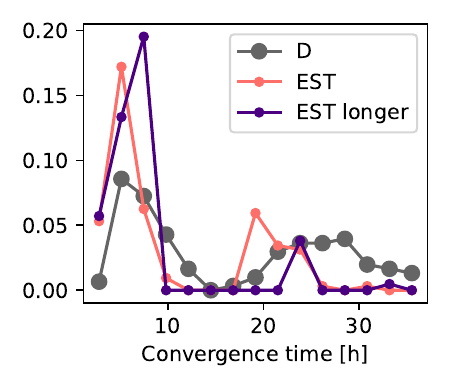}}
\subfigure[$\eta=0.9$]{\includegraphics[width=0.24\textwidth]{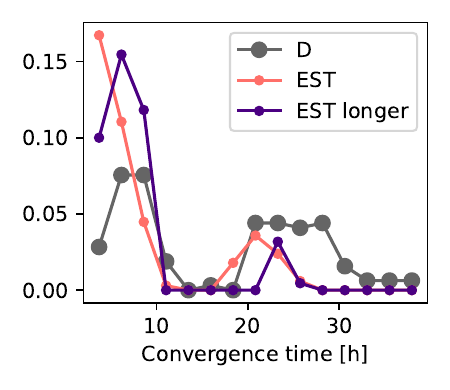}}
\subfigure[$\eta=1$]{\includegraphics[width=0.24\textwidth]{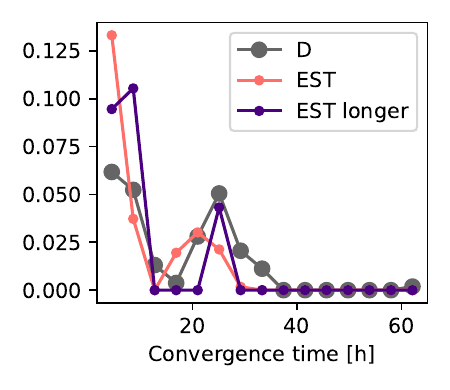}}
\caption{\textbf{Naming game with longer temporal length}. The results with longer surrogates (EST longer), lasting 40 days, are compared with those of the EST surrogates of 2 days (EST) and with the primary school dataset (D).}
\label{fig_NG_longer}
\end{figure*}

The method that we propose can generate surrogates with a longer time span than the original networks. 
In the main text we only showed surrogates with the same temporal length of the original network, in order to compare them and assess the quality of the surrogates.
We show here some results obtained by generating longer networks (40 days instead of the 2 days of the primary school) and simulating the dynamical processes on them, see fig.~\ref{fig_SIR_longer} for the SIR model and fig.\ref{fig_NG_longer} for the naming game (the Deffuant model is not affected by the temporal length because the process is much faster than the others and always converges before reaching the end of the dataset).
We observe that the general results are very similar to those obtained with the shorter surrogates and the dataset (both repeated multiple times to reach the stationary states of the dynamics). 

\section{Temporal states and hour partition}

\begin{figure}
\includegraphics[width=0.5\textwidth]{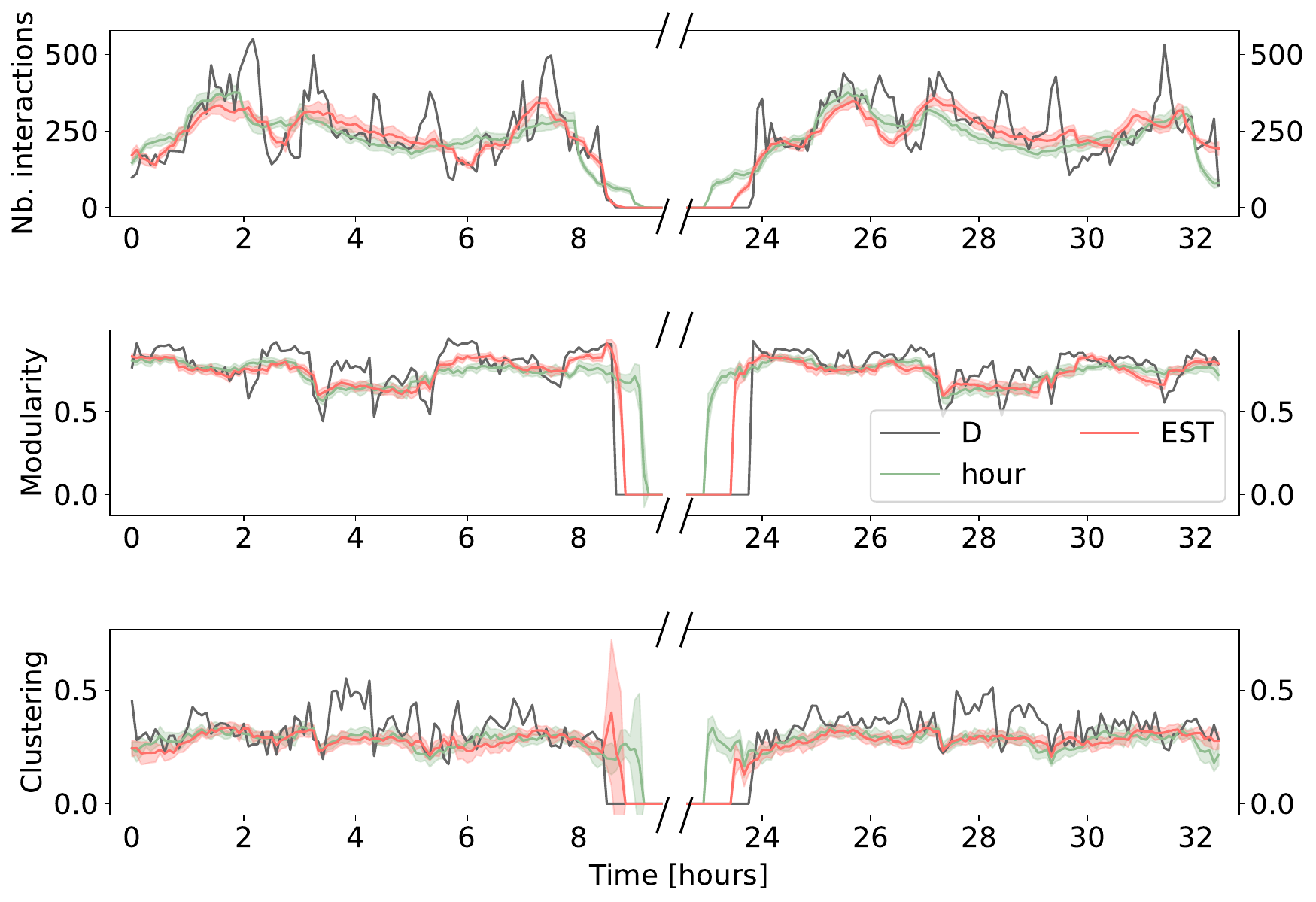}
\includegraphics[width=0.5\textwidth]{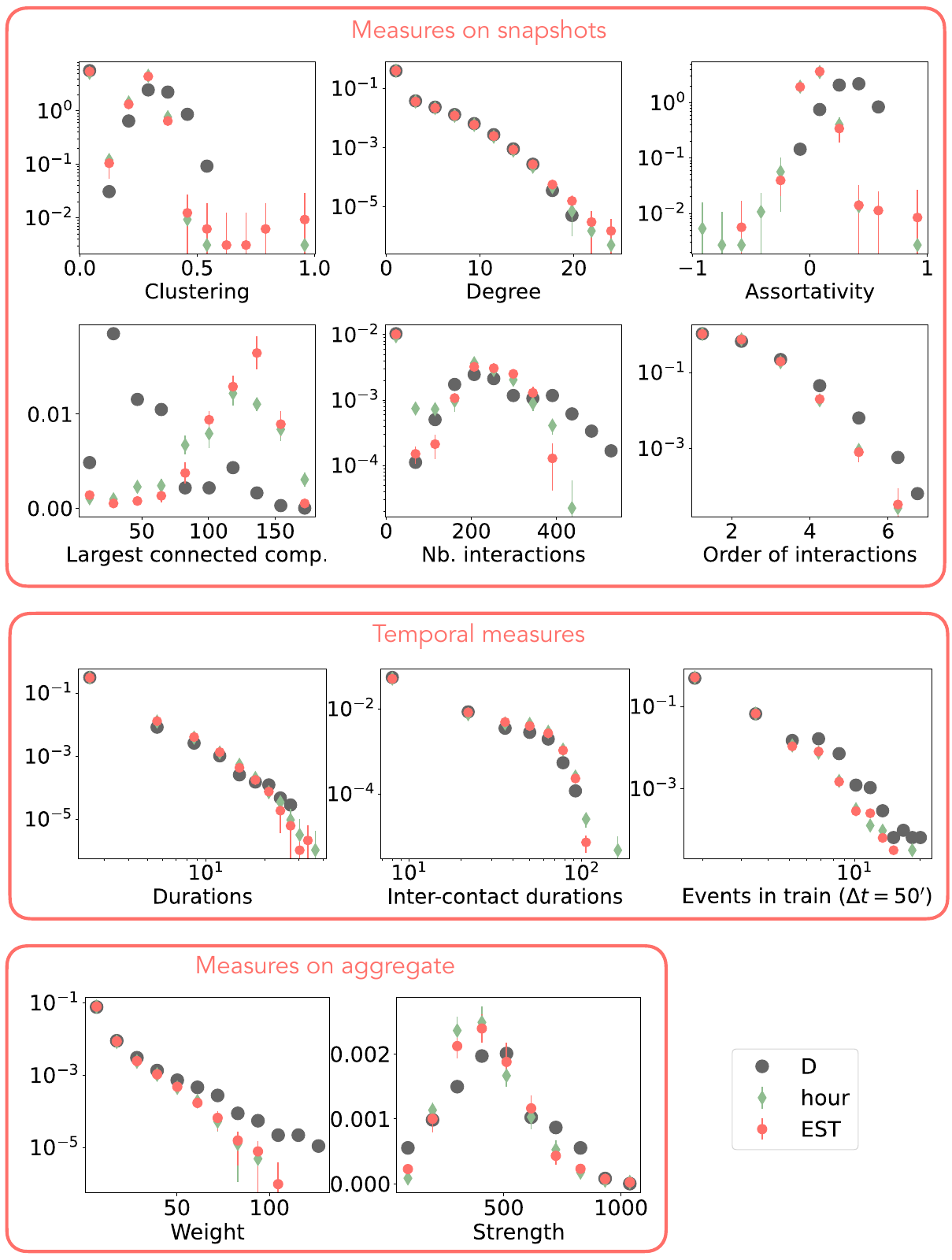}
\caption{\textbf{Temporal states vs. hour partition}. Structural and temporal measures for the primary school dataset (D), the surrogate networks obtained with the EST procedure and the surrogates obtained with a time partition into hours.} 
\label{fig_hour}
\end{figure}

The proposed algorithm implies dividing the time span of the surrogate network in temporal states ($s_e$ and $s_{\mathcal{L}}$) to assign different characteristics that are observed to change in time in the original networks.
The original idea based on Egocentric Temporal Neighborhoods, proposed by Longa et al.\cite{longa2024generating},  was instead based on a partition of the original network into slots of one hour arbitrarily chosen.
However this partition is not very general, while the partition into states  reflects better the real temporal behavior of a dataset and can be applied to more general contexts.
In fig.~\ref{fig_hour} we compare the surrogates obtained with the partition of  with the partition into hours of the day. 
The similarity of the results found with the two methods in fig.~\ref{fig_hour} proves that the partition into states is not less accurate than that into hours, while being more versatile.

\section{Triangles intra and inter communities}

\begin{figure}
\includegraphics[width=0.5\textwidth]{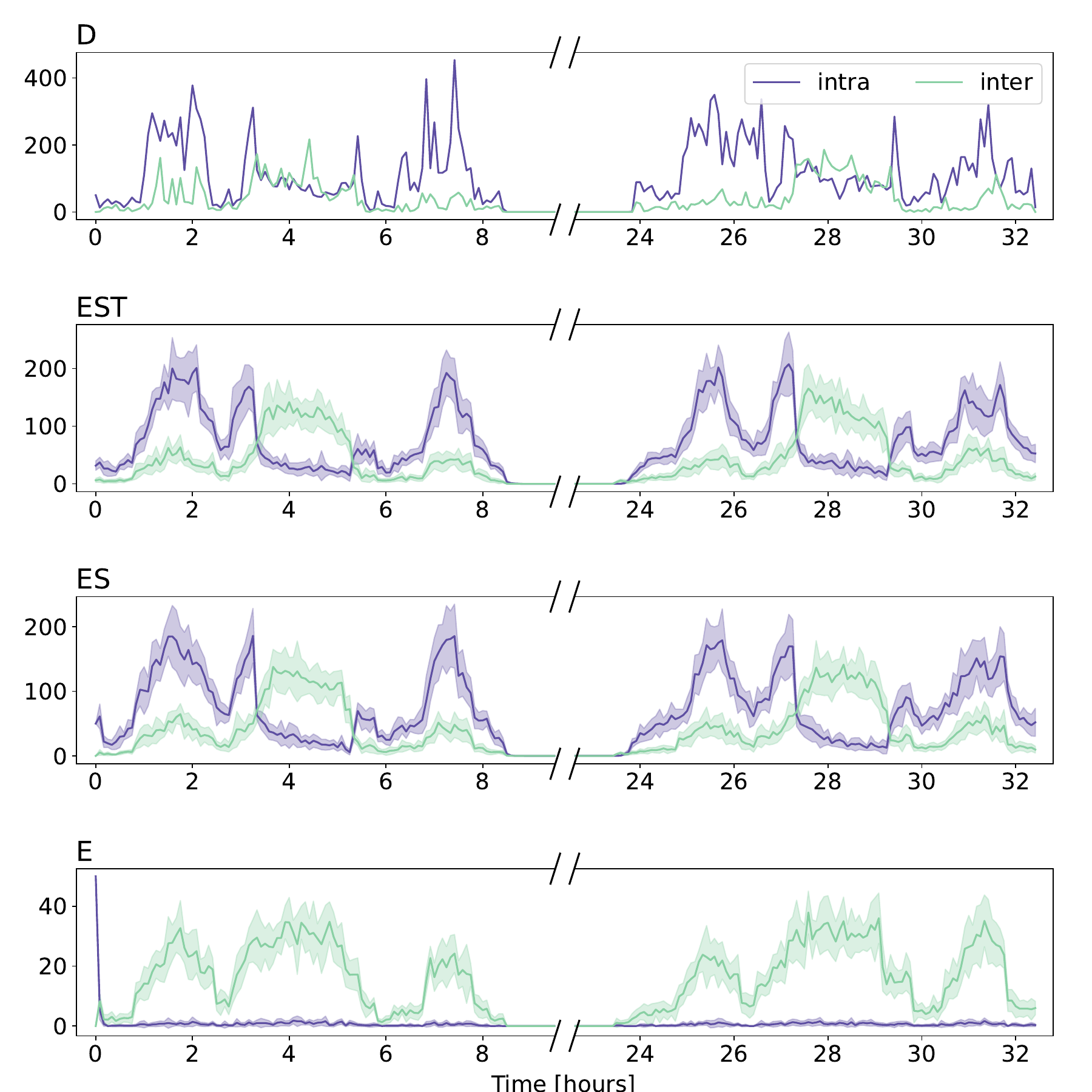}
\caption{\textbf{Number of closed triangles intra and inter communities}. The dataset is the primary school and the communities are given by metadata.}
\label{fig_tri_intra_inter}
\end{figure}

In the main text we have measured the clustering (i.e. the number of closed triangles with respect to the number of triads, or open triangles) of each layer of the surrogate temporal networks, comparing it to the datasets.
We here investigate the role of communities in the appearance of closed triangles.
In fig.~\ref{fig_tri_intra_inter} we report, for the primary school dataset, the number of triangles whose nodes are part of the same community (denoted as "intra") and the number of triangles whose nodes belong to at least two different communities (denoted as "inter").
The communities here are defined by the school classes provided by metadata.
We notice that in the dataset the triangles take place more often inside communities, especially during the classes, while during the lunch breaks the number of triangles intra and inter communities become similar.
If a surrogate is generated without structure (E) only a few triangles appear and they are almost all across communities.
This is trivially due to the higher number of possible links that connect nodes belonging to different communities.
We recover a realistic ratio when introducing the structural features (with the ES and EST methods) that show a higher number of intra-communities triangles during classes. 
Both methods however overestimate the number of inter-communities triangles during lunch breaks.


\section{Additional datasets}

\subsection{High school 11}

\begin{figure*}
\includegraphics[width=\textwidth]{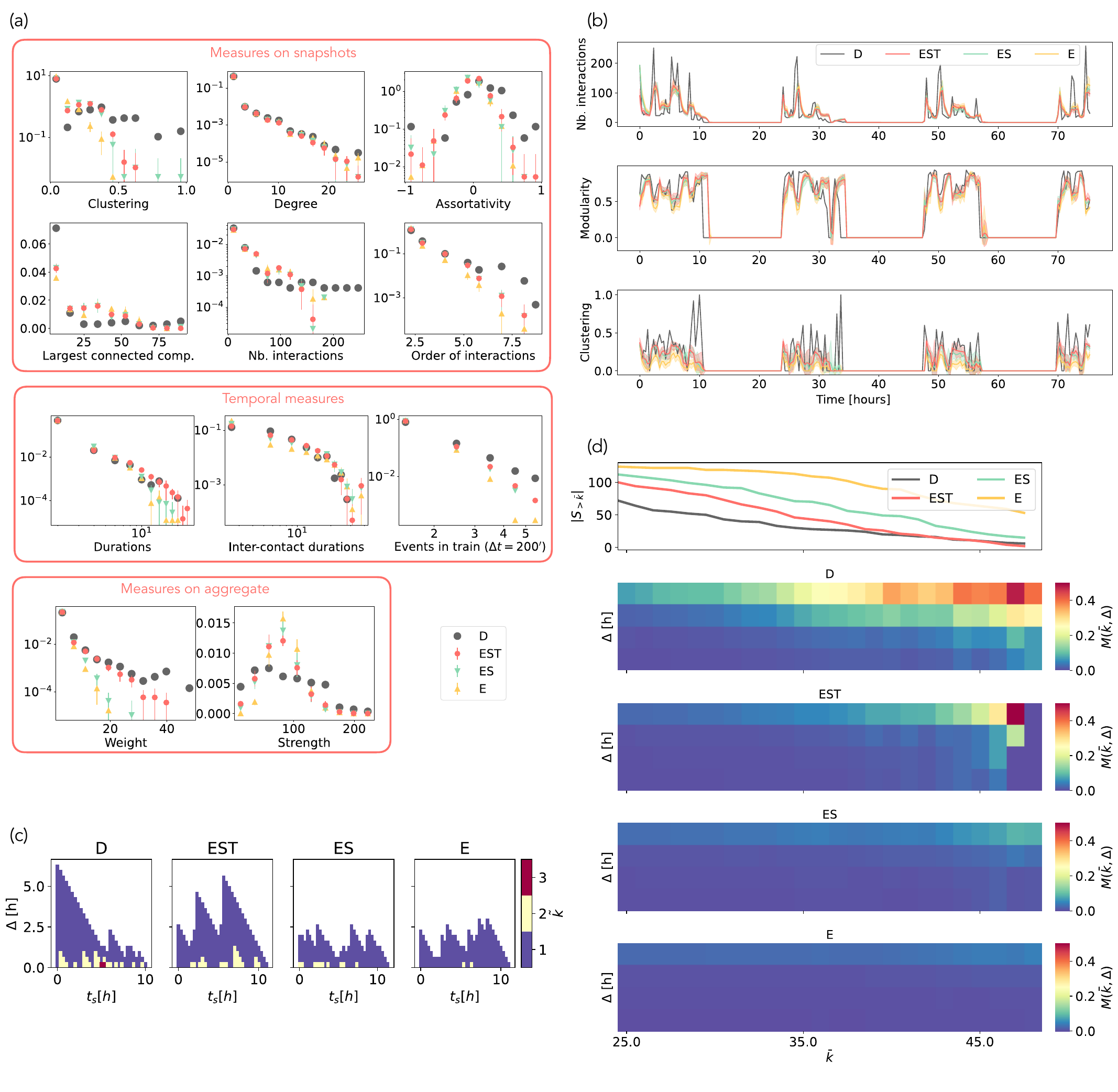}
\caption{Structural and temporal properties of high school 11 dataset and surrogates.}
\label{fig_top_hschool11}
\end{figure*}

The high school 11 dataset~\cite{fournet2014contact} describes the contacts among 126 kids and teachers in a high school for 4 days.
The surrogate networks that we obtain for this dataset are analyzed and compared to the original network on their structural and temporal properties, as reported in fig.~\ref{fig_top_hschool11}.

\begin{figure*}
\includegraphics[width=\textwidth]{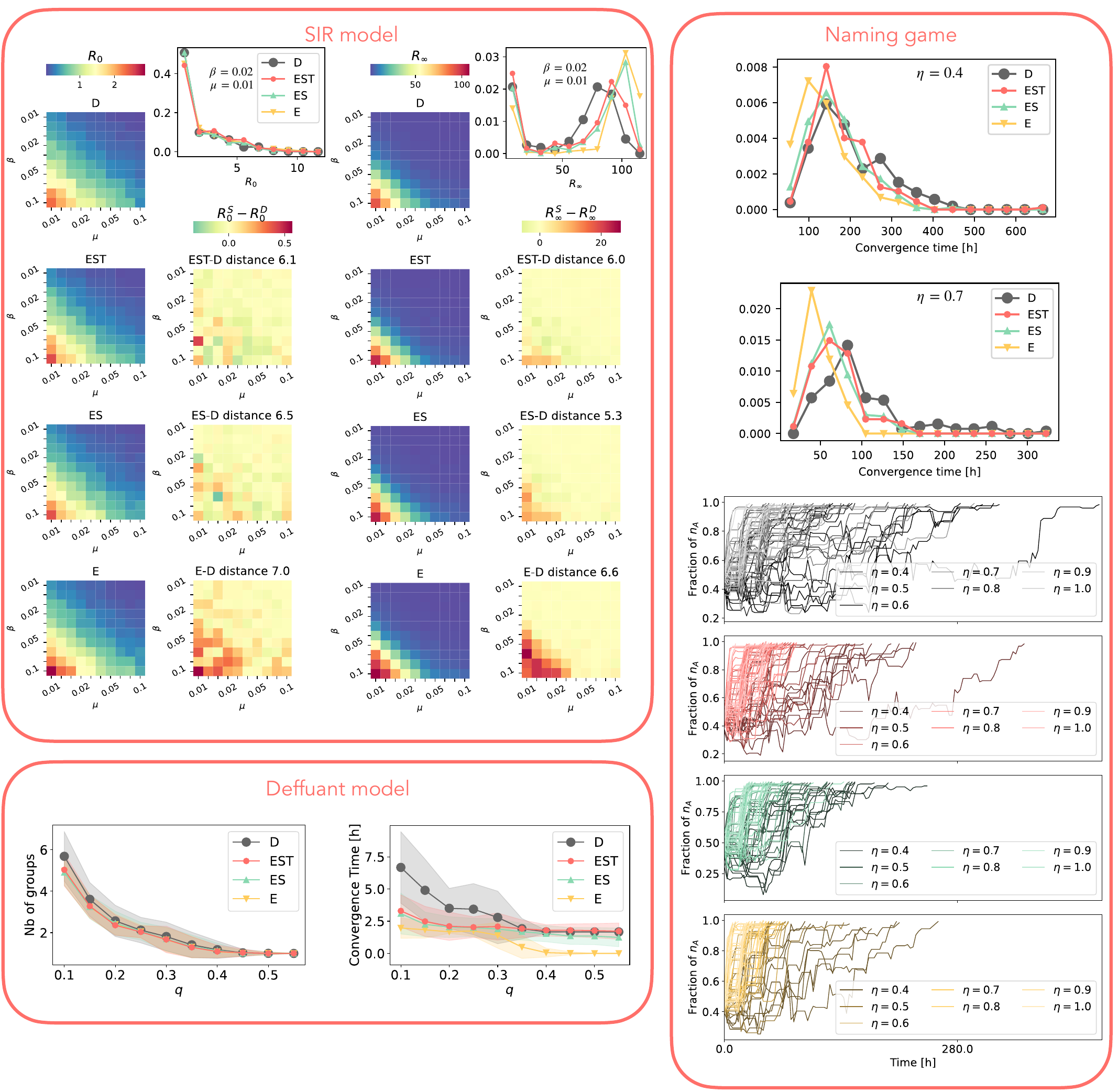}
\caption{Dynamical processes on high school 11 dataset and surrogates.}
\label{fig_dyn_hschool11}
\end{figure*}

The results of the dynamical processes, SIR model, Deffuant model, and Naming game, are shown in fig.~\ref{fig_dyn_hschool11}.

\subsection{High school 13}
\begin{figure*}
\includegraphics[width=\textwidth]{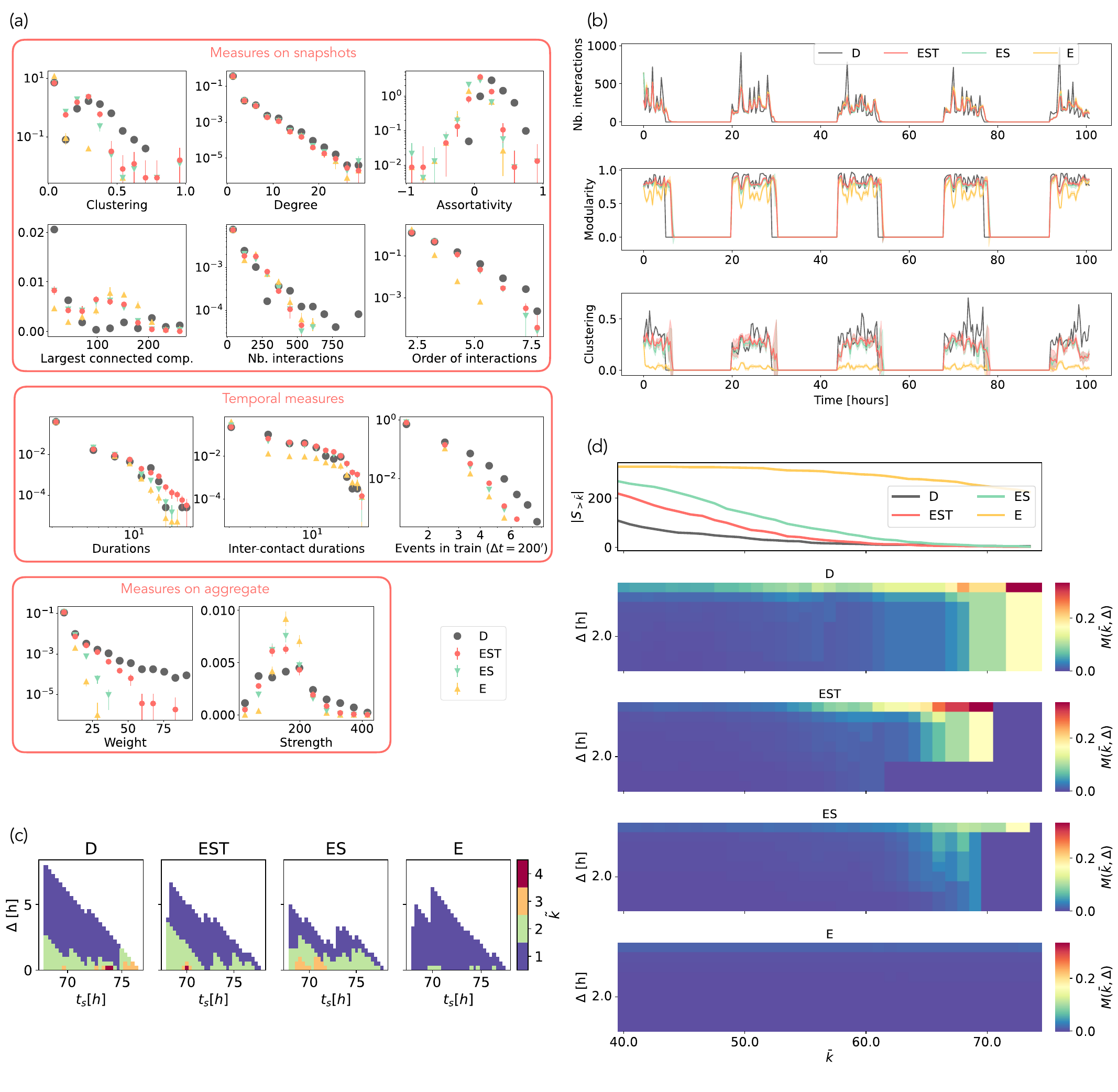}
\caption{Structural and temporal properties of high school 13 dataset and surrogates.}
\label{fig_top_hschool13}
\end{figure*}

The high school 13 dataset~\cite{mastrandrea2015contact} describes the contacts among 327 kids and teachers in a high school for 5 days.
The surrogate networks that we obtain for this dataset are analyzed and compared to the original network on their structural and temporal properties, as reported in fig.~\ref{fig_top_hschool13}.

\begin{figure*}
\includegraphics[width=\textwidth]{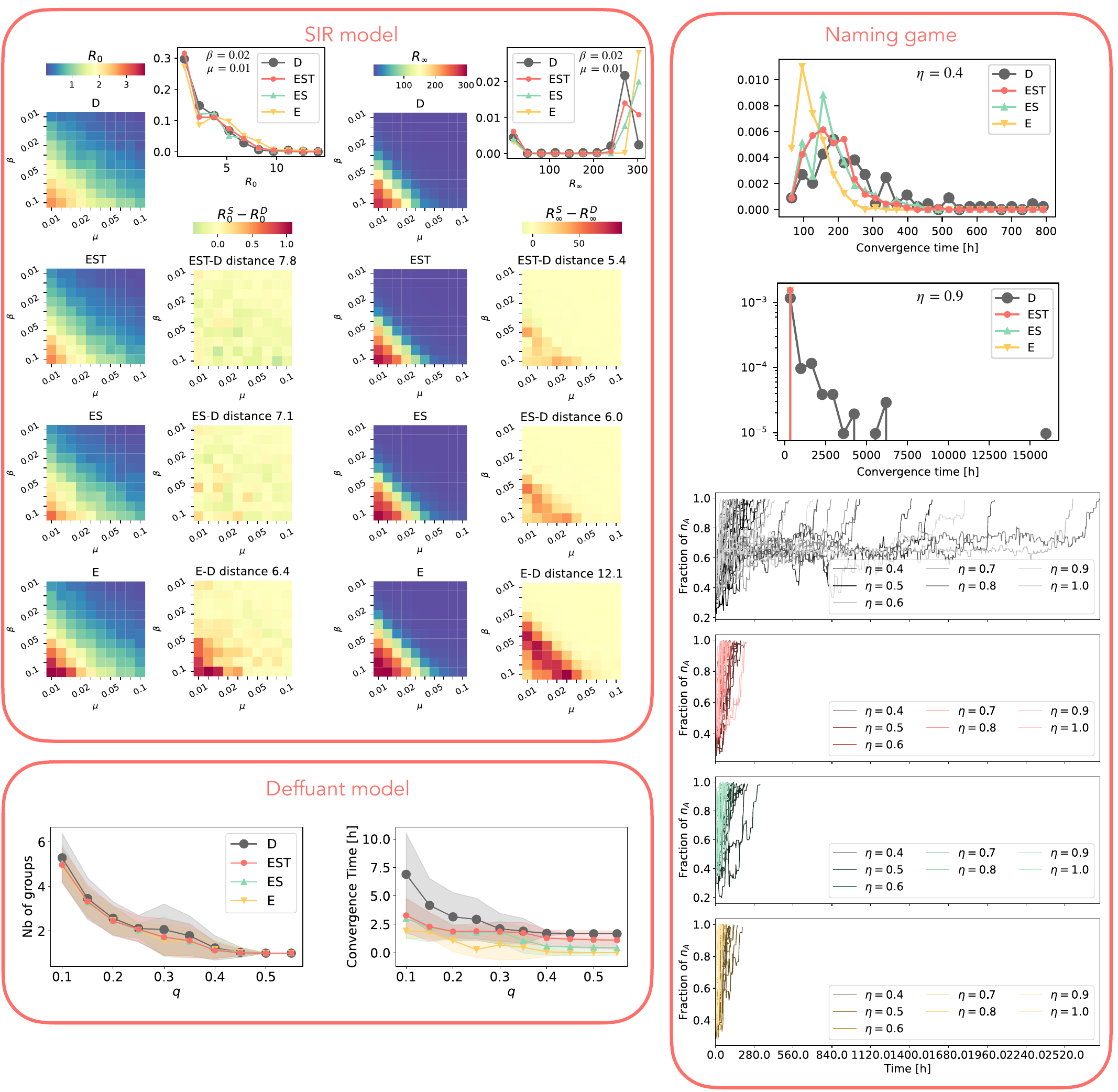}
\caption{Dynamical processes on high school 13 dataset and surrogates.}
\label{fig_dyn_hschool13}
\end{figure*}

The results of the dynamical processes, SIR model, Deffuant model, and Naming game, are shown in fig.~\ref{fig_dyn_hschool13}.

For the Naming Game we observe that the process taking place on the original network takes a very long time to converge, a peculiarity that is not observed in the other datasets and that is not reproduced by the surrogates.
This happens especially for high values of $\eta$ and is partially due to the fact that, as mentioned in section~\ref{SI_sec_hmod}, the high school 13 dataset is characterized by a hierarchical modularity that the surrogates are not able to reproduce with the standard generation method.
In fact, the modularity that is reproduced by the surrogates does not consider that the inter-communities links are not distributed among all the communities but are concentrated on some of them, leading in this example to two groups of communities (one with classes \textit{2BIO1}, \textit{2BIO2}, and \textit{2BIO3}, and one with \textit{MP}, \textit{MP*1}, \textit{MP*2}, \textit{PC}, \textit{PC*}, and \textit{PSI*}).
This, when the Naming Game is implemented, can imply a high polarization of these two groups by the two possible names $A$ and $B$: for high values of $\eta$ the nodes inside a group easily agree on the same name and, if the other group compactly agrees on the other name, the probability that one group overcomes the other one by convincing all the members of it is very low. 
It hence takes a long time before the entire population finds an agreement.

\begin{figure}
\subfigure[Naming game]{\includegraphics[width=0.49\textwidth]{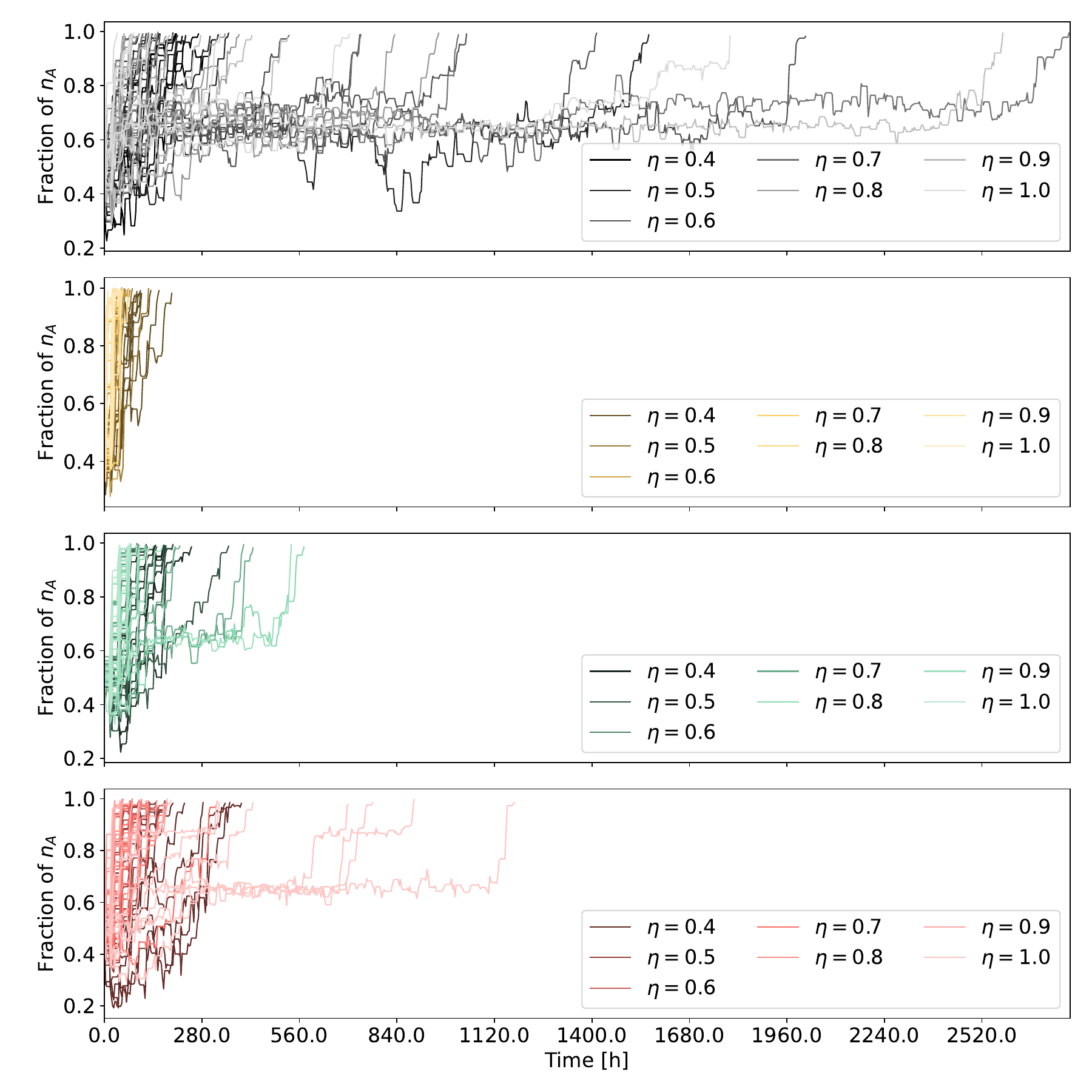}}
\caption{Naming game high school 13 where the EST and ES surrogates have been obtained with hierarchical modularity.}
\label{fig_ng_hschool13_SBM}
\end{figure}

This effect is partially recovered with the surrogate networks obtained with the method for hierarchical modularity explained in section~\ref{SI_sec_hmod}: the results of the process simulated on these networks are reported in fig.~\ref{fig_ng_hschool13_SBM} (where only the EST and ES results are different from the Naming game panel in fig.~\ref{fig_dyn_hschool13}). 
For some simulations  we indeed observe a very long convergence time, even if we are still far from the large distribution obtained for the original network.

\subsection{Conference}

\begin{figure*}
\includegraphics[width=\textwidth]{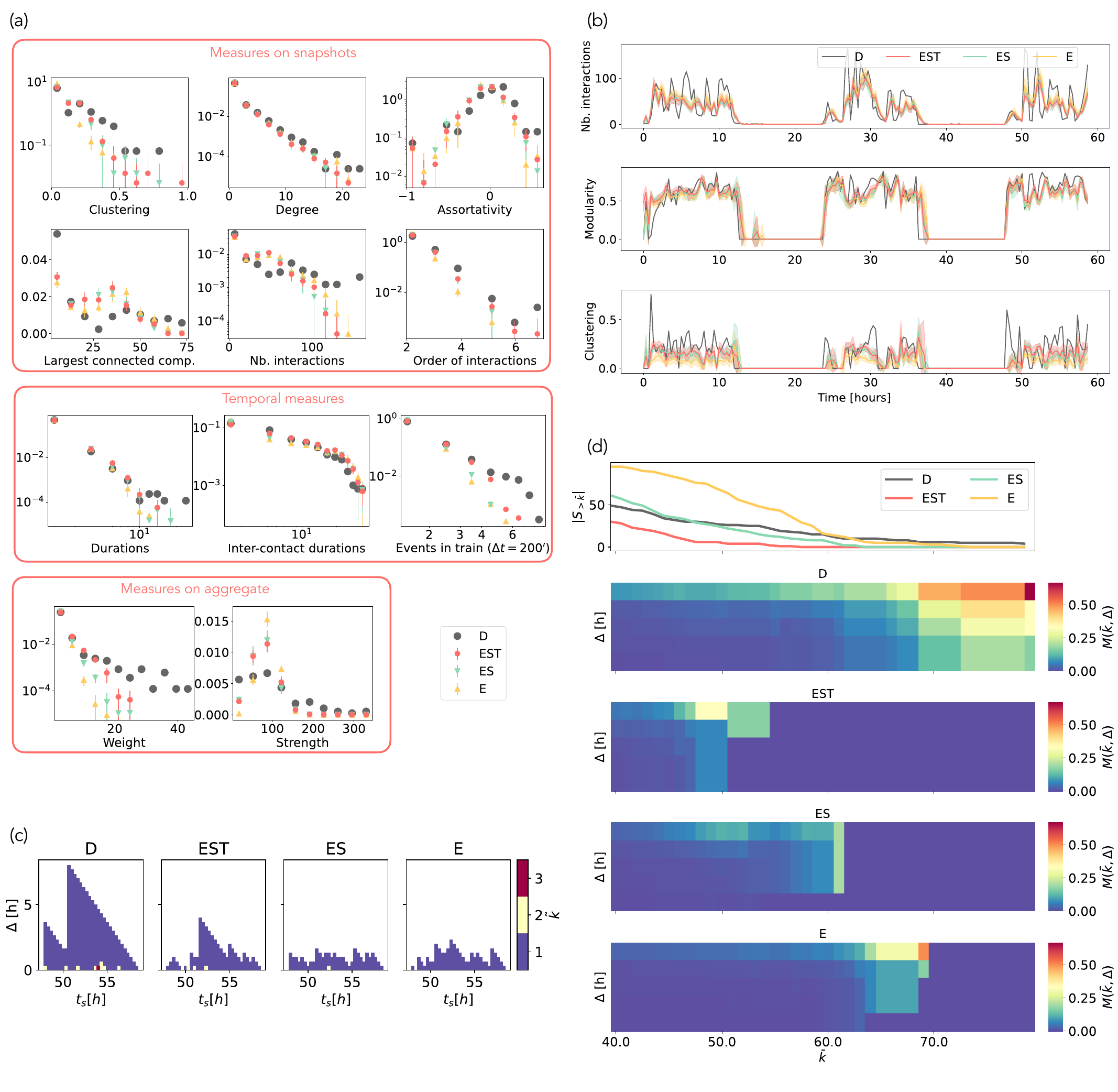}
\caption{Structural and temporal properties of conference dataset and surrogates.}
\label{fig_top_hypertext}
\end{figure*}

The conference dataset~\cite{isella2011s} describes the contacts among 113 participants to a conference for 3 days.
The surrogate networks that we obtain for this dataset are analyzed and compared to the original network on their structural and temporal properties, as reported in fig.~\ref{fig_top_hypertext}.

\begin{figure*}
\includegraphics[width=\textwidth]{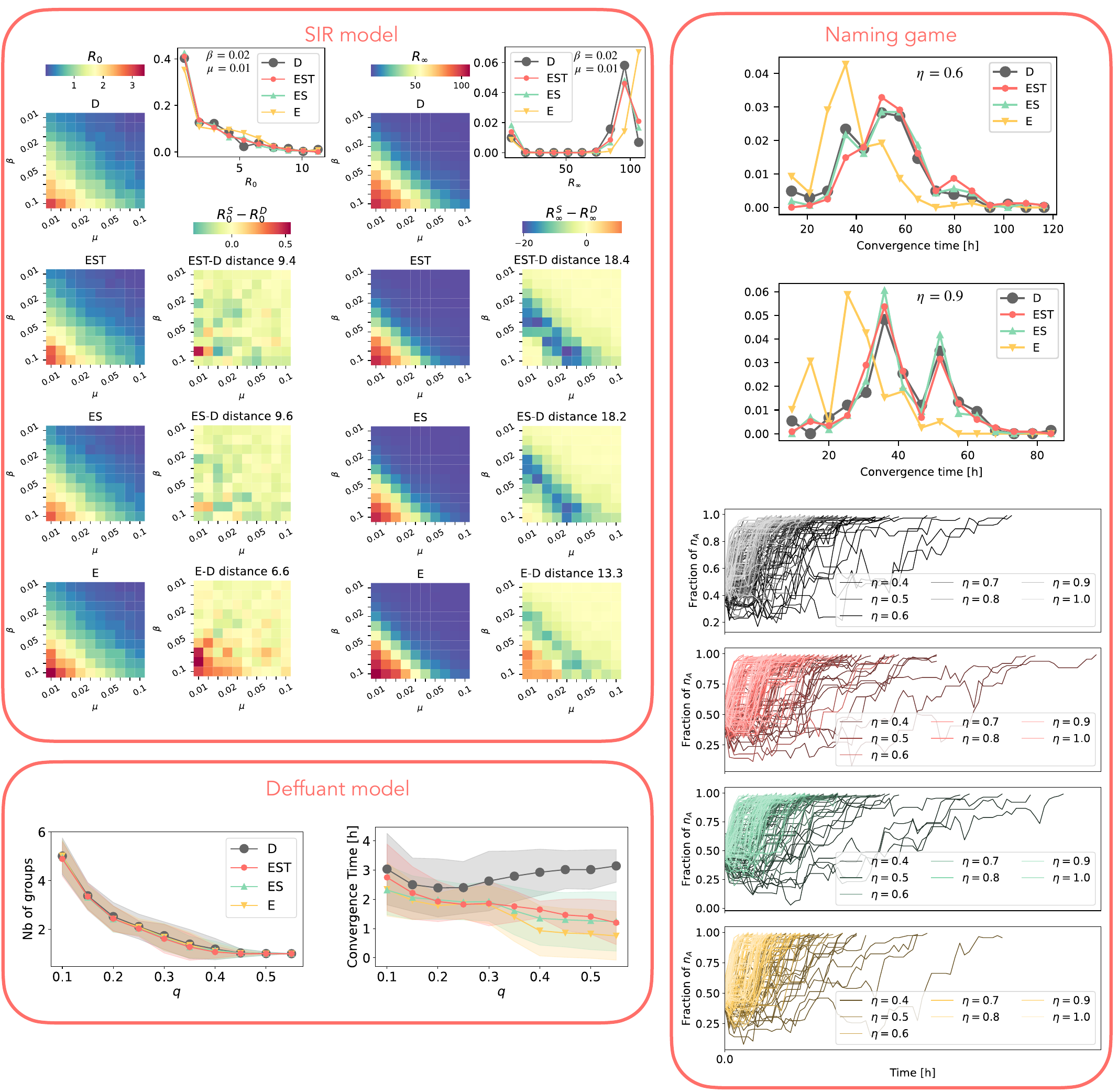}
\caption{Dynamical processes on conference dataset and surrogates.}
\label{fig_dyn_hypertext}
\end{figure*}

The results of the dynamical processes, SIR model, Deffuant model, and Naming game, are shown in fig.~\ref{fig_dyn_hypertext}.

\subsection{Workplace}

\begin{figure*}
\includegraphics[width=\textwidth]{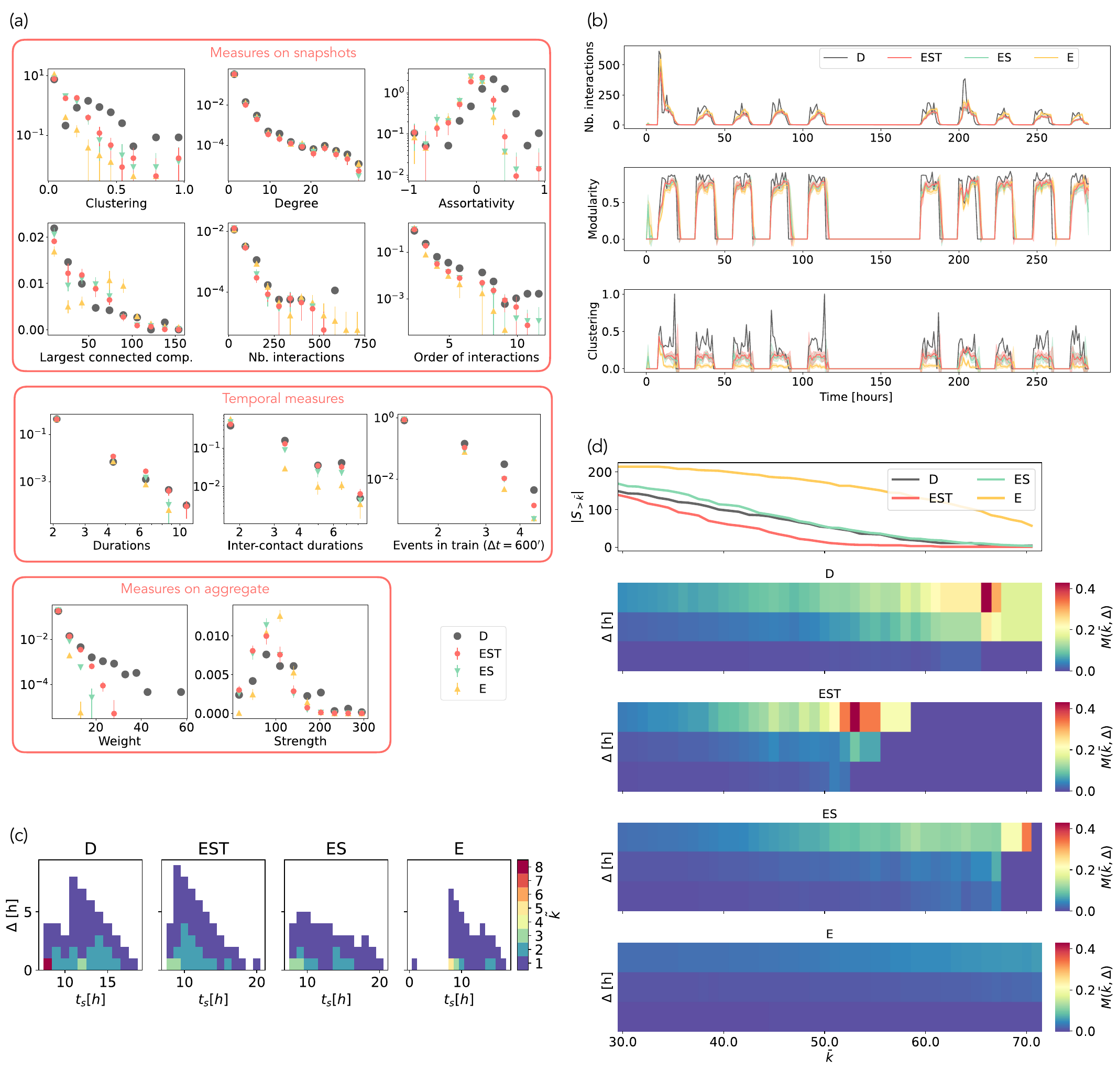}
\caption{Structural and temporal properties of workplace dataset and surrogates.}
\label{fig_top_InVS15}
\end{figure*}

The workplace dataset~\cite{genois2018can} describes the contacts among 217 employees for 12 days.
The surrogate networks that we obtain for this dataset are analyzed and compared to the original network on their structural and temporal properties, as reported in figs.~\ref{fig_top_InVS15}.

\begin{figure*}
\includegraphics[width=\textwidth]{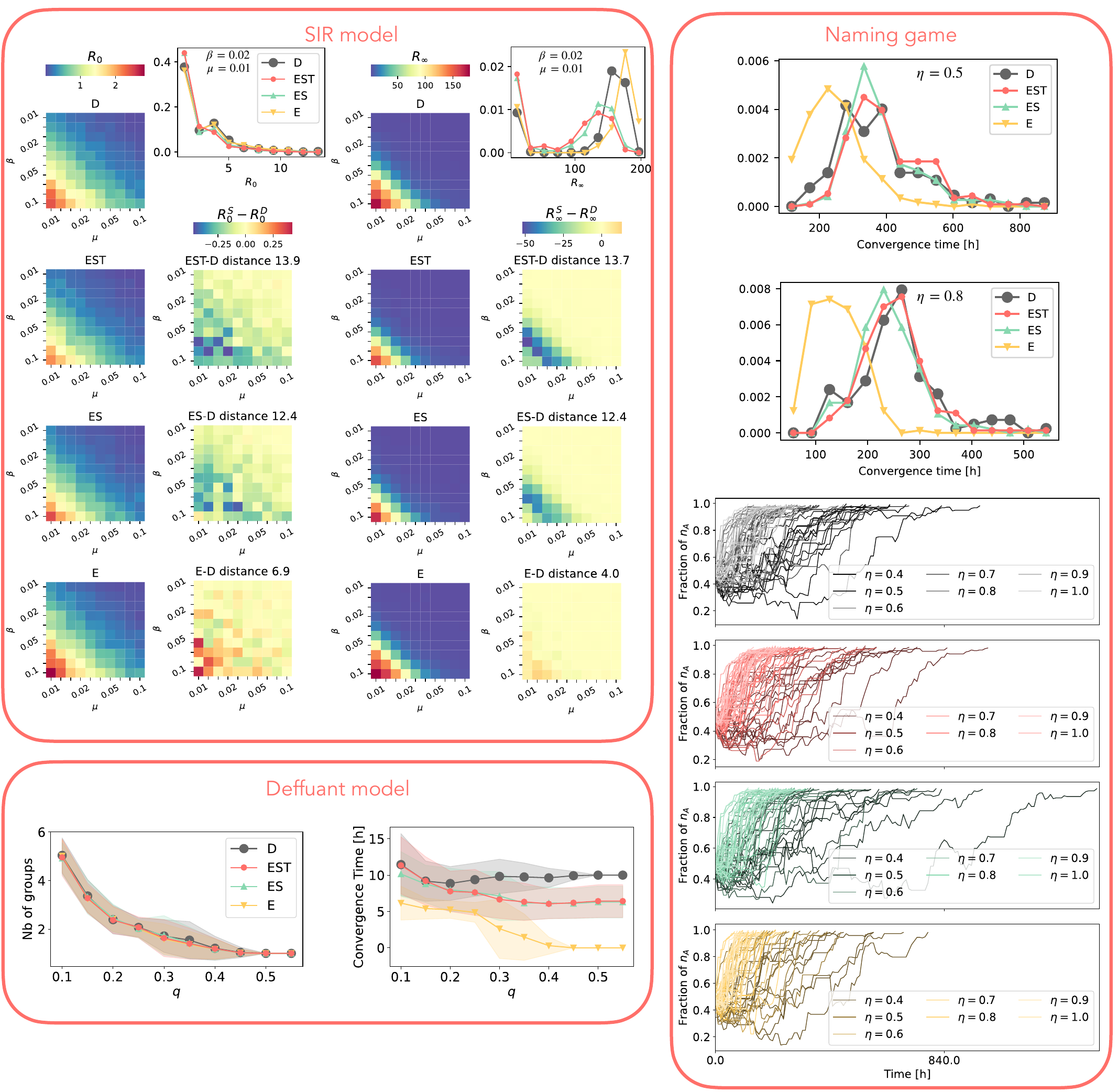}
\caption{Dynamical processes on workplace dataset and surrogates.}
\label{fig_dyn_InVS15}
\end{figure*}

The results of the dynamical processes, SIR model, Deffuant model, and Naming game, are shown in fig.~\ref{fig_dyn_InVS15}.

\section{Epidemic threshold}

\begin{figure*}
\subfigure[Primary school]{\includegraphics[width=0.3\textwidth]{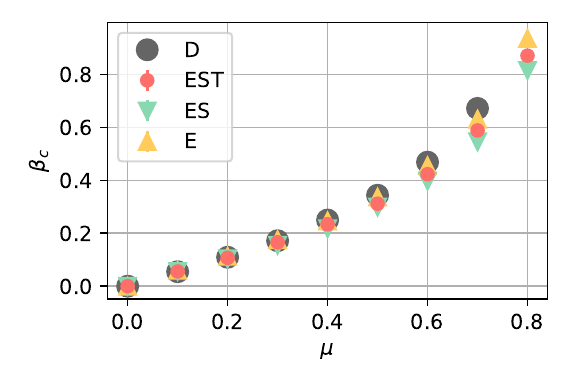}} 
\subfigure[High school 11]{\includegraphics[width=0.3\textwidth]{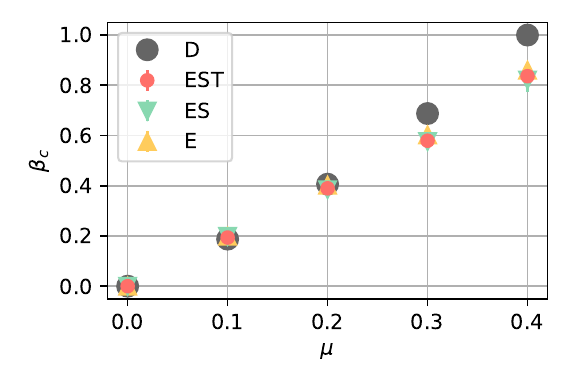}} 
\subfigure[High school 13]{\includegraphics[width=0.3\textwidth]{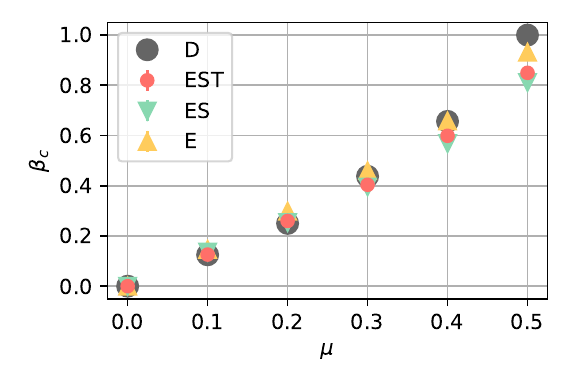}}
\subfigure[Conference]{\includegraphics[width=0.3\textwidth]{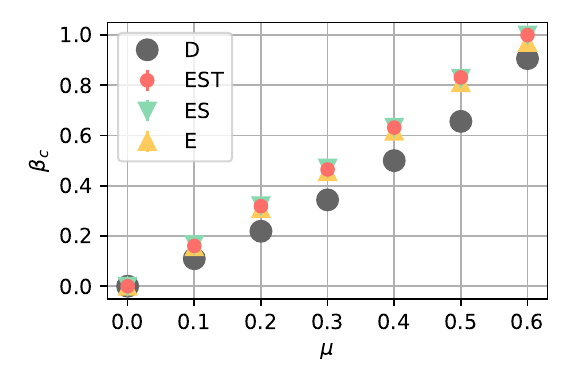}}
\subfigure[Workplace]{\includegraphics[width=0.3\textwidth]{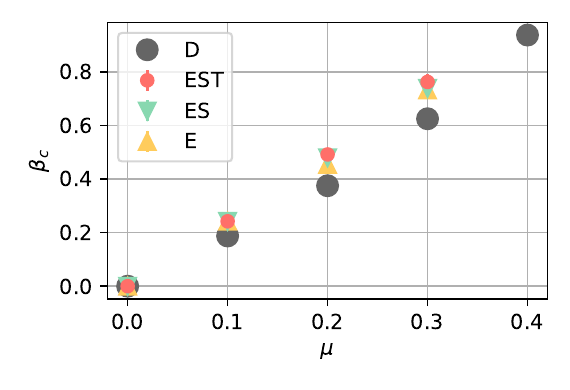}} 
\caption{Epidemic threshold.}
\label{fig_epi}
\end{figure*}

We have assessed the surrogates' quality by analysing the results of three dynamical processes simulated on them. 
We here propose an additional measure that quantifies the spreading efficiency of a temporal network: the epidemic threshold, which corresponds to the critical condition for the widespreading regime.
We in particular compute, given a value of recovery rate $\mu$, the critical value of infection rate $\beta_c$.
The computation can be performed analytically by using the method proposed by Valdano et al.~\cite{valdano2015analytical,valdano2015infection} and is reported in fig.~\ref{fig_epi}.

%

\end{document}